\documentclass[usenatbib,usegraphicx,twocolumn]{mn2e}

\def\ie{{\it i.e. }}
\def\eg{{\it e.g. }}

\def\simlt{\mathrel{\hbox{\rlap{\hbox{\lower4pt\hbox{$\sim$}}}\hbox{$<$}}}}
\def\simgt{\mathrel{\hbox{\rlap{\hbox{\lower4pt\hbox{$\sim$}}}\hbox{$>$}}}}

\begin{document}

\title[The WFCAM Photometric System] {The UKIRT Wide Field
  Camera $Z$$Y$$J$$H$$K$ Photometric System: Calibration from 2MASS}

\author[Hodgkin et al.]
{S. T. Hodgkin$^{1}$, M. J. Irwin$^{1}$, P. C. Hewett$^{1}$,
  S. J. Warren$^{2}$\\
$^1$Institute of Astronomy, Madingley Road, Cambridge CB3 0HA\\
$^2$Astrophysics Group, Blackett Laboratory, Imperial College London, Prince 
Consort Road, London SW7 2BW
}

\date{Accepted
      Received
      in original form}

\maketitle

\begin{abstract}

  In this paper we describe the photometric calibration of data taken
  with the near-infrared Wide Field Camera (WFCAM) on the United
  Kingdom Infrared Telescope (UKIRT). The broadband $Z$$Y$$J$$H$$K$
  data are directly calibrated from 2MASS point sources which are
  abundant in every WFCAM pointing. We perform an analysis of spatial
  systematics in the photometric calibration, both inter- and
  intra-detector and show that these are present at up to the
  $\sim$5\,per cent level in WFCAM. Although the causes of these
  systematics are not yet fully understood, a method for their removal
  is developed and tested. Following application of the correction
  procedure the photometric calibration of WFCAM is found to be
  accurate to $\simeq$1.5\,per cent for the $J$$H$$K$ bands and 2\,per
  cent for the $Z$$Y$ bands, meeting the survey requirements. We
  investigate the transformations between the 2MASS and WFCAM systems
  and find that the $Z$ and $Y$ calibration is sensitive to the
  effects of interstellar reddening for large values of $E(B-V)'$, but
  that the $J$$H$$K$ filters remain largely unaffected. We measure a
  small correction to the WFCAM $Y$-band photometry required to place
  WFCAM on a Vega system, and investigate WFCAM measurements of
  published standard stars from the list of UKIRT faint
  standards. Finally we present empirically determined throughput
  measurements for WFCAM.

\end{abstract}


\section{Introduction} 


This paper describes the photometric calibration of the United Kingdom
Infrared Telescope (UKIRT) Wide Field Camera (WFCAM)
$Z$$Y$$J$$H$$K$ passbands spanning $0.84-2.37\,\mu$m. The WFCAM
photometric system, based on synthetic colours, is described in Hewett
et al. (2006).

WFCAM is currently mounted on UKIRT about 75\,per cent of the time, the
majority of which is used to perform a set of five public surveys,
under the umbrella of the UKIRT Infrared Deep Sky Survey: UKIDSS,
(Lawrence et al. 2007, Warren et al. 2007), as well as a number of
campaign projects and other smaller programmes.

All WFCAM data are processed and calibrated by the Cambridge
Astronomical Survey Unit (CASU) using an automated pipeline (Irwin et
al. 2008). Due to the large data rate of WFCAM (150--230 Gigabytes of
raw image data are taken per night), a reliable and robust method for
photometric calibration is required. The primary photometric
calibrators are drawn from stars in the Two Micron all Sky Survey
(2MASS, Skrutskie et al. 2006), present in large numbers in every
WFCAM exposure (the median number is around 200 per pointing). The
2MASS calibration has been shown to be uniform across the whole sky to
better than 2\,per cent accuracy (Nikolaev et al. 2000).

In this paper we investigate the accuracy and homogeneity of the WFCAM
calibration, by analysis of data taken in the first two years of
science operations. The survey requirement is to provide a photometric
calibration in broad band filters accurate to 2\,per cent with a goal
of 1\,per cent. To test whether the calibrated photometry passes the
requirements, we consider the following tests:

\begin{itemize}
\item for repeat observations of non-variable stars (of sufficient
  signal-to-noise), the measured $rms$ (root mean square) should be no
  worse than 0.02 magnitudes.
\item the calibration should be robust against the effects of
  reddening, i.e. the calibrated frame zeropoints, measured on
  photometric nights, should have a standard deviation $\leq 0.02$
  magnitudes as a function of reddening.
\item the photometric calibration should be uniform across the sky,
  and not depend on the observed stellar population or the number of
  stars in the image. Specifically, the calibrated frame zeropoints
  (on photometric nights) should have a 1$\sigma$ scatter of $\leq
  0.02$ magnitudes even in sparse regions of sky, e.g. at high
  Galactic latitude.
\end{itemize}

We demonstrate that for the vast majority of observations with WFCAM,
these tests are passed, and we quantify the conditions under which
they are not.

The paper is organised as follows: Section~\ref{method} summarises how
the WFCAM data are calibrated within the pipeline; in
Section~\ref{repeatability} we examine the repeatability of WFCAM
photometry, quantify residual spatial systematics, and investigate the
limits of the calibration in non-photometric conditions; in
Section~\ref{2mcomparison} we investigate the relationship between the
2MASS and WFCAM photometric systems and investigate the effects of
interstellar reddening; in Section~\ref{bias} we investigate the
extent of bias in the WFCAM calibration arising from the
overestimation of flux in faint sources in 2MASS; in
Section~\ref{secoffsets} we compare the WFCAM system to a {\it Vega}
system where A0 stars should have zero colour in all passbands; in
Section~\ref{ukirtfs} we compare the WFCAM photometry with published
data for faint standards measured with UKIRT; Finally, in
Section~\ref{throughput} we derive the throughput of WFCAM. At the
time of writing, UKIDSS has completed its fourth Data Release (DR4).

 
%
%
%
%
%
%
%














\section{Method for routine photometric calibration of WFCAM
  data}\label{method}

All WFCAM data are processed in Cambridge with a pipeline developed by
CASU and documented in Irwin et al. (2008). In this section we
describe in detail the steps used to measure magnitudes and then apply
a photometric calibration for each WFCAM image and subsequent
catalogue. After processing and calibration, the WFCAM data (in the
form of FITS binary images and catalogues) are transferred to
Edinburgh for ingestion into the WFCAM Science Archive (WSA, Hambly
et al. 2008). The UKIDSS data have now seen a number of data releases
from the WSA, and the photometric calibration has evolved to tackle
the (increasingly smaller) corrections needed to meet the survey
goals; the corrections included in each release are summarised in
Table~\ref{tab:releases}.

\begin{table}
\centering
\caption{Summary of photometric calibration corrections included in
  each UKIDSS release from the WSA. The columns indicate which of the
  following are accounted for: a reddening 
  dependent correction (see Section~\ref{sec:galext}), a per-chip
  zeropoint (discussed in 
  Section~\ref{photcal}), a radial distortion term
  (Section~\ref{instru}), and a residual 2D spatial systematic
  correction (Section~\ref{ressys}).}
\begin{tabular}{lrrrrr}\hline
Release & Date      & E(B-V)' & chip ZP & radial & spatial\\
\hline
EDR     & Feb 2006  &  no &  no &  no &  no\\
DR1     & Jul 2006  &  no &  no &  no &  no\\
DR2     & Mar 2007  & yes &  no &  no &  no\\
DR3     & Dec 2007  & yes & yes &  no &  no\\
DR4     & Jul 2008  & yes & yes & yes & yes\\ 
\hline
\end{tabular}
\label{tab:releases}
\end{table}

\subsection{Key WFCAM elements}

WFCAM comprises four Rockwell Hawaii-II PACE arrays, with 2k $\times$
2k pixels at 0.4 arcsec/pixel, giving a solid angle of 0.21 deg$^2$
per exposure. A detailed description of WFCAM can be found in Casali
et al. (2007). A prerequisite for the calibration scheme is to ensure
that all four detectors are calibrated to the same photometric
system. The implementation of the reduction pipeline needs to ensure
that the different intrinsic detector quantum efficiencies (and
possibly their different colour dependence) are correctly taken into
account. Thus much of the calibration is performed {\it per-detector}.

There are five broadband filters designed for WFCAM, the
$Z$$Y$$J$$H$$K$ filters, which are described in detail by Hewett et
al. (2006) and characterized using synthetic photometry. The $J$$H$$K$
filters possess transmission profiles specified to reproduce as
closely as possible the widely adopted MKO system (Tokunaga, Simons \&
Vacca 2005). The $Z$ and $Y$ filters cover 0.84--0.93$\mu$m and
0.97--1.07$\mu$m respectively.

\subsection{Measurement of instrumental magnitudes}\label{instru}

The source extraction software (Irwin et al. 2008) measures an array
of background-subtracted aperture fluxes for each detected source,
using 13 soft-edged circular apertures of radius $r/2$, $r/\sqrt{2}$,
$r$, $\sqrt{2} r$, $2r$ ... up to $12r$, where $r=1$ arcsecond. A
soft-edged aperture divides the flux in pixels lying across the
aperture boundary in proportion to the pixel area enclosed. In this
paper we only consider photometry derived from fluxes measured within
an aperture of radius 1 arcsecond. However, all the apertures of
selected isolated bright stars are used to determine the
curve-of-growth of the aperture fluxes, i.e. the enclosed counts as a
function of radius. This curve of growth is used to measure the point
spread function (PSF) aperture correction for point sources for each
detector, for each aperture (up to and including $4r$, which includes
typically $\sim$99\,per cent, or more, of the total stellar
flux). Irwin et al. (2008) find that this method derives aperture
corrections which contribute $\leq1$\,per cent to the overall
photometry error budget


A further correction should be applied to the source flux to account
for the non-negligible field distortion in WFCAM, described in detail
in Irwin et al. (2008). The astrometric distortion is radial and leads
to an increase in pixel area by around 1.2\,per cent compared to the
centre of the field of view. Standard image processing techniques
assume a uniform pixel scale, and that a correctly reduced image will
have a flat background. For WFCAM's variable pixel scale, this is
actually incorrect and one would expect to see an increase in the sky
counts per pixel at large off-axis angles, while the total number of
counts detected from a star would be independent of its position on
the array. The flatfielding of an image therefore introduces a
systematic error into the photometry of sources towards the edge of
the field of view. The corrected flux $f_{\rm cor}$,
where $f$ is defined as the aperture corrected count-rate in ADU per
second of the source above background, is simply:

\begin{equation}
f_{\rm cor} = f / ( 1 + 3 k_3 r^2 ) ( 1 + k_3 r^2 )
\label{eq:distortcor}
\end{equation}

Where $k_3$ (with units of radian/radian$^3$) is the coefficient of
the third order polynomial term in the radial distortion equation
(Irwin et al. 2008) and is found to be slowly wavelength dependent
with a value of -50.0 in the $H$-band (called PV2\_3 in the FITS
headers). The instrumental magnitude is then 

\begin{equation}
m_i=-2.5log_{10}(f_{\rm cor})
\end{equation}

\begin{figure}
\begin{center}
  \includegraphics[width=8.5cm,viewport=25 00 570
  540,clip]{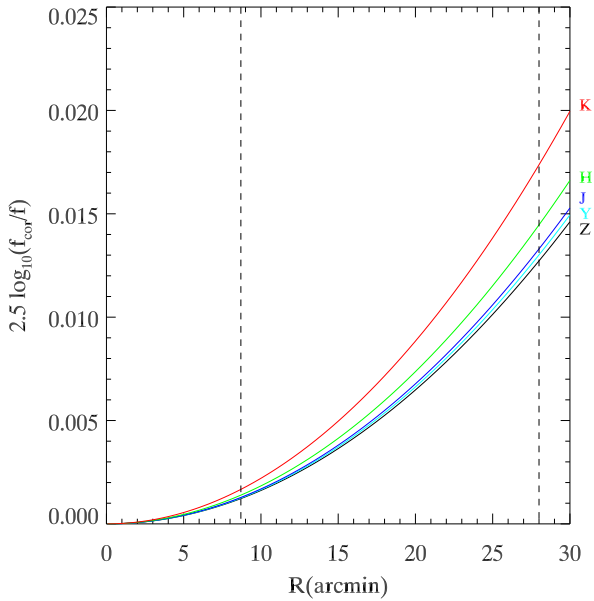}
  \caption{Photometric radial distortion (magnitudes) in WFCAM as a
    function of offaxis angle (arcmin). There are curves for all five
    broadband filters, $Z$$Y$$J$$H$$K$ from bottom to top. The
    WFCAM detectors are separated by 94\,per cent of their active area
    in a square pattern (see Fig.~\ref{figillum}), thus the vertical
    lines represent the location of the detector corners.}
\label{figraddis}
\end{center}
\end{figure}

This correction has not been applied for WSA releases DR1--DR3 (Hambly
et al. 2008), but is included for DR4 and subsequent
releases. Fig.~\ref{figraddis} plots the radial distortion term
($f_{\rm cor}/f$ converted to magnitudes) as a function of off-axis
angle for the WFCAM filters.

\subsection{Calibration of the photometry}\label{photcal}

The data are firstly flatfielded using twilight flatfields (which are
updated monthly), and an initial gain correction is applied to place
all four detectors on a common system, to first order. The
per-detector magnitude ZP is then derived for each frame from
measurements of stars in the 2MASS point source catalogue (PSC) that
fall within the same frame. Thus the calibration stars are measured at
the same time as the target sources, taking us away from more
traditional calibration schemes, whereby standard star observations
are interspersed with target observations.

We assume that there exists a simple linear relation between the
stellar 2MASS and WFCAM colours, e.g. $J_{\rm w}-J_2 \propto
J_2-H_2$. In a Vega-based photometric system, this relation should
pass through (0,0), i.e. for an A0 star
$Z$\,=\,$Y$\,=\,$J$\,=\,$H$\,=\,$K$, irrespective of the filter system
in use. For each star in 2MASS observed with WFCAM, the pipeline
derives a ZP (at airmass unity) from

\begin{equation}
ZP  = m_{\rm 2} +  C (J_2-H_2\, {\rm or}\, J_2-K_2) - m_i + k(\chi-1) 
\end{equation}

where $m_i$ is the aperture corrected instrumental magnitude (derived
above). $C$ are the colour coefficients for each passband and have
been solved for by combining data from many nights (see
Section~\ref{2mcomparison}). $k$ is a default value for the extinction
(0.05 in all filters, see below) and $\chi$ is the airmass.

The 2MASS sources used for the calibration of WFCAM are selected to
have extinction-corrected colour $0.0 \le J_2-K_2 \le 1.0$ with a
2MASS signal-to-noise ratio $>10$ in each filter. If fewer than 25
2MASS sources are found within the field-of-view of the detector, then
the colour cut is not applied. The WFCAM astrometric calibration is
also derived from 2MASS and for both systems the astrometric $rms$
accuracy per source is $<$ 100mas (Irwin et al. 2008). The maximum
allowed separation for a 2MASS-WFCAM source match is 1 arcsecond. For
all WFCAM observations up until 2008 March 28th, the median number of
2MASS stars falling within a WFCAM detector, which match our colour
and signal-to-noise requirements, is $\sim 200$. Only around 0.1\,per
cent of WFCAM pointings have fewer than 25 2MASS stars covering a
single detector. For the UKIDSS Large Area Survey (LAS) alone, the
median number of 2MASS standards per detector is $\sim 100$, and
again, around 0.1\,per cent have fewer than 25.

For a single pointing, for each detector, the zeropoint is then
derived as the median of all the per-star zeropoint values. A single
photometric zeropoint for the pointing, MAGZPT, is calculated as the
median of the detector zeropoints over all 4 detectors. The associated
error, MAGZRR, is 1.48$\times$ the median absolute deviation of the
detector zeropoints around the median. This error therefore comprises
several components: the intrinsic errors in the 2MASS photometry, the
error in the conversion from 2MASS to the WFCAM filter, and then any
residual systematic offsets between the detectors (see
Section~\ref{d2doff}). In the $Z$-band, the MAGZRR errors are
the largest, typically about twice that for the $J$$H$$K$
filters.

It should be noted that we do not derive any atmospheric extinction
terms on a given night with WFCAM. Rather, the value of MAGZPT derived
above incorporates an instantaneous measure of extinction at the
observed airmass. The photometric calibration of a field therefore
includes no error from this assumption. However the derived zeropoint
for airmass unity will include a small error (because we assume the
extinction is 0.05 magnitudes per airmass for all filters). Leggett et
al. (2006) find the extinction at Mauna Kea to be $k_J=0.047$,
$k_H=0.029$, $k_K=0.052$ with standard deviations between 0.2 and
0.3. For a typical WFCAM frame, observed at an airmass$\approx 1.3$,
an extinction which differs from our assumed value by 0.03
magnitude/airmass will lead to a 0.01 magnitude error in the value of
MAGZPT. The value of MAGZPT over time can be used to investigate the
long term sensitivity of WFCAM due to, for example, the accumulation
of dust on the optical surfaces, and seasonal variations in
extinction.

\subsection{Detector offsets}

A final stage to the photometric calibration takes account of
systematic differences between the four detectors, measured on a
monthly basis. The object catalogues associated with each science
product frame are used in the pipeline calibration process to compute
a single per pointing overall zeropoint for the array of 4 detectors,
using the colour equations specified in Section~\ref{2mcomparison}.
The residuals from all 2MASS stars used in the frame zeropoint
determination (\ie $J$, $H$, $K$ signal:noise $>$ 10:1) are also
computed on a per pointing basis together with their standard
coordinate location ($\xi,\eta$) with respect to the tangent point of
the telescope optical axis (see e.g. Irwin et al. 2008). We use
standard coordinates since these are independent of the degree of
interleaving and dithering, and hence pixel scales, used in
constructing the science product images and catalogues.  These
residuals are then partitioned by month and stacked. The monthly
timescale corresponds with the changover of the master flat field
frames, with which we anticipate some of the corrections are
correlated. From UKIDSS DR3, all WFCAM photometric calibration is
determined and applied monthly. The details of this correction process
are discussed in Section~\ref{spatialmeas}. The result is to generate
zeropoints per detector (the value of MAGZPT is updated for each
detector).

The pipeline also estimates the nightly zeropoint (NIGHTZPT) and an
associated error (NIGHTZRR) which can be used to gauge the
photometricity of a night. NIGHTZPT is simply the median of all ZPs
measured within the night, and NIGHTZRR is a measure of the scatter in
NIGHTZPT.

\section{Repeatability of the photometry}\label{repeatability}

\subsection{Overlap analysis from the UKIDSS Large Area Survey}\label{repeatlas}

\begin{figure}
\begin{center}
  \includegraphics[width=17.0cm,viewport=20 00 800
  550,clip]{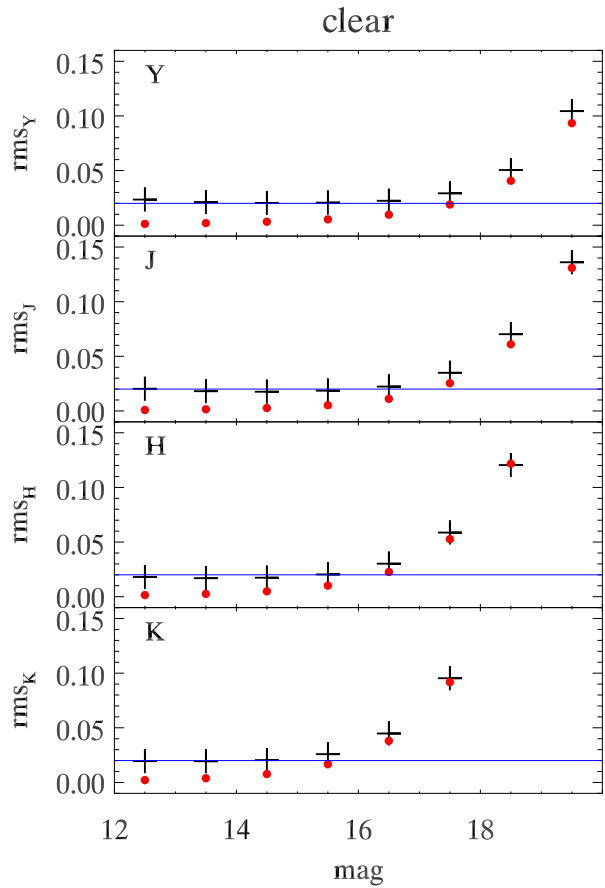}
  \caption{$rms$ diagram for repeat observations of sources in the DR3
    dataset for the LAS (selected to be stellar in both observations)
    for the $Y$$J$$H$$K$ filters (top to bottom), showing results for
    data taken in photometric conditions. Photometric conditions are
    defined as having a zeropoint within 0.05 magnitudes of the
    survey median value. Black crosses are 1.48 $\times$ the MAD
    (median absolute deviation) of the data in magnitude slices,
    i.e. equivalent to the Gaussian sigma of the magnitude bin. Red
    circles are the median pipeline errors for sources in the bin as
    reported in the WFCAM Science Archive for DR3. The blue horizontal
    lines at 0.02 magnitudes represent the target 2\,per cent
    photometric reliability.}
\label{lasdr3repeat}
\end{center}
\end{figure}

We can use repeat observations of sources in UKIDSS to investigate the
photometric accuracy and homogeneity of the survey and to compare with
the pipeline errors.  The pipeline errors are calculated from the
source counts and local background only, and do not include
corrections for bad pixels and other detector artifacts. In addition,
they deliberately do not include any contribution from the calibration
procedure itself (i.e. the inherent uncertainties in the calibrators)
or additional corrections for residual systematic effects in the
calibration. By comparing multiple observations of sources near the
edges of the detectors we are more sensitive to systematics from
spatial effects, e.g. scattered light in the flatfield, a variable
point spread function, and so on (see Section~\ref{spatial}). Hence
this analysis demonstrates the worst case, and should help put upper
limits on the error contribution from systematic contributions.
 
Fig.~\ref{lasdr3repeat} shows the $rms$ diagrams for repeat
measurements, for all point sources (classed as stellar in both
observations) in the DR3 release of the LAS. Repeat observations arise
from the small overlap of detector edges between the four pointings
which go towards constructing a tile, as well as from observations of
adjacent tiles.  The $rms$ for each magnitude bin is calculated from
the standard deviation of the differences between the two measurements
from all the repeats, $\sigma_{\Delta mag}/\sqrt 2$. The data used in
Fig.~\ref{lasdr3repeat} were observed in photometric conditions, with
a frame photometric zeropoint within 0.05 magnitudes of the median
value for all observations made with the filter ($\sigma_{\rm
  ZP}=0.03$ magnitudes in all pass bands). We exclude sources at the
very edges of the array, within 10 arcseconds of the detector edges.

For a photometric calibration good to 2\,per cent, then the median
magnitude difference should be zero, and the $rms$ of the distribution
for the brighter stars should be $\sigma/\sqrt 2 = 0.02$. 

\begin{figure}
\centering
\includegraphics[width=8.5cm,viewport=20 10 520 390,clip]{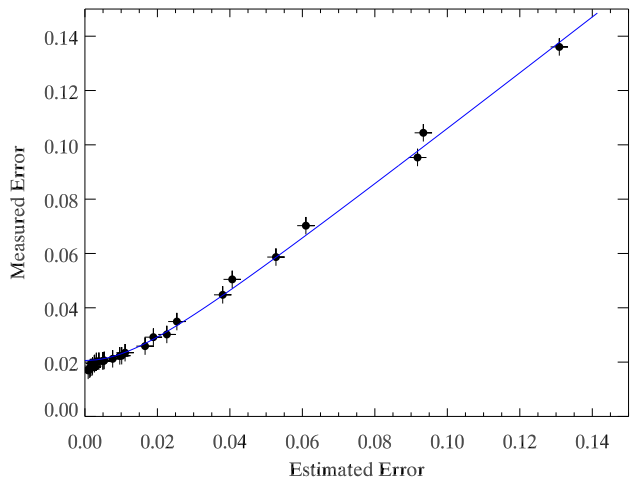}
\caption{A plot of measured errors ($M$) against estimated errors
  ($E$) derived from the $rms$ distributions for LAS overlaps. The
  solid line is a linear fit to $M^2$ versus $E^2$ and is described in
  the text.}
\label{errorserrors}
\end{figure}

At the bright end, the $rms$ (black crosses in
Fig.~\ref{lasdr3repeat}) is very close to the survey goal of 2\,per
cent. Unsurprisingly, the pipeline estimated random noise errors (red
dots in Fig.~\ref{lasdr3repeat}) significantly underpredict the
observed distribution at the bright end, indicating that systematic
errors, dominate here at the 2\,per cent
level. Figure~\ref{errorserrors} combines measurements for all filters
and compares the estimated random photon noise errors, $E$, derived by
the pipeline to the measured errors $M$ (the $rms$ described above),
which include both systematic calibration errors and random components
due to photon noise. The points have been fitted with a simple
relation of the form $ M^2 = c E^2 + s^2 $ where the systematic
component $s=0.021 \pm 0.001$, the constant of proportionality
$c=1.082 \pm 0.014$.  In principle these can be used to update the
default pipeline error estimates, however, we suspect that $c$ is
slightly greater than unity due to a combination of factors relating
to edge effects on the detectors and noise covariance arising from
inter-pixel capacitance within the detector (Irwin et al. 2008), which
the pipeline error estimates do not allow for.

\subsection{Spatial systematics}\label{spatial}


\subsubsection{Measurement}\label{spatialmeas}

As described in Section~\ref{photcal}, we store the standard
coordinates ($\xi,\eta$) and WFCAM magnitude for each observation of a
2MASS star. For each month of data and for each filter we then compute
a per star zeropoint relative to the previously computed overall
zeropoint for each science product image for the entire array.  These
differences are then stacked within a fixed standard coordinate grid,
of cell size $\approx 1.2 \times 1.2$ arcmin covering the complete
array of four detectors, making use of selected data from each month of
observations.  Only science products taken in photometric conditions
are used in this process to minimise the effects of non-photometric
residual structure.  The average offsets in the grid of values are
then filtered (smoothed) using a combination of a two-dimensional
3-pixel bimedian and bilinear filter.  This latter step ensures a smooth
variation of values over the grid by effectively correlating the
corrections on a scale of neighbouring grid points. This is necessary
since, even with a month of data, the number of points per cell is
still only typically 25-100.  The $rms$ noise per cell from the 2MASS
errors alone is $\approx$1\,per cent.  With
smoothing, this noise drops to the few milli-mag level which is a
negligible extra error with respect to the derived corrections.

\begin{figure}
\centering
\includegraphics[width=8.0cm,angle=270]{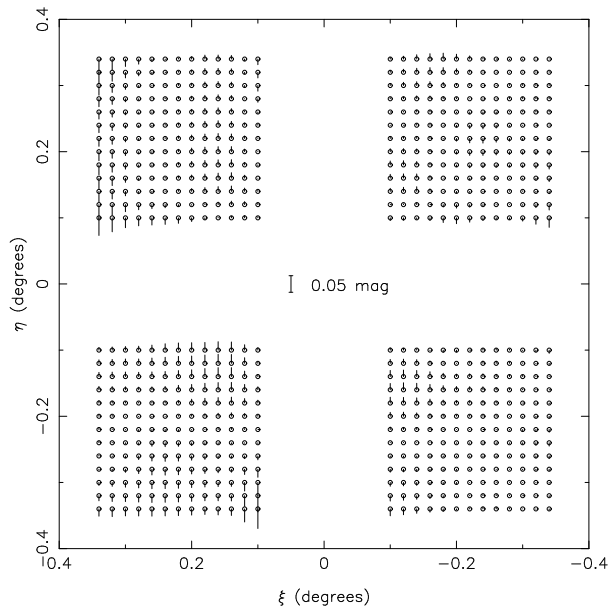}
\caption{An example of the remaining monthly average spatial magnitude 
residuals for WFCAM $J$-band observations in 2005 September after applying 
passband-dependent individual detector zeropoint corrections.  The residuals
are shown in standard coordinates with respect to the tangent point of
the observations, i.e. E is to to the left and N is to the top.}
\label{figillum}
\end{figure}

The residuals for each detector are then grouped and the median
correction calculated. These median factors define the detector-level 
zeropoint
corrections for each passband.  The final stage is to apply the
detector-level corrections and compute the residual systematics which
are recorded as a correction table, and diagnostic plot.  An example
of the latter is shown in Fig.~\ref{figillum} and the possible origins of
these patterns are discussed below. The correction tables are
currently available via the CASU web pages\footnote{http://casu.ast.cam.ac.uk/surveys-projects/wfcam/data-processing/illumination-corrections}.

\subsubsection{Detector-to-detector offsets}\label{d2doff}

\begin{figure}
\centering
\includegraphics[width=8.5cm]{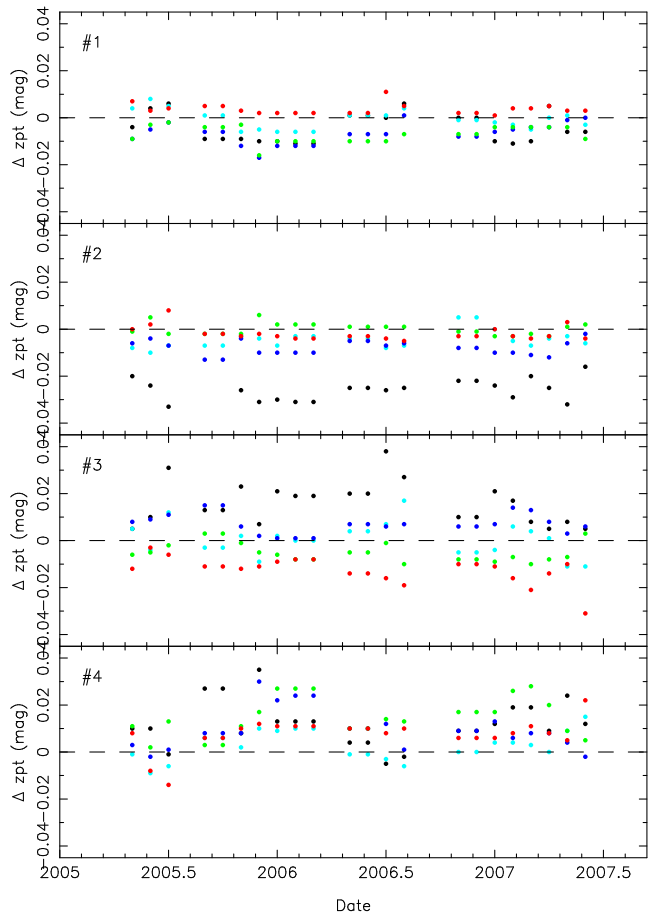}
\caption{An overview of the computed detector zeropoint corrections as
  a function of observation date for all of the WFCAM broadband
  filters: $Z$ - black; $Y$ - cyan; $J$ - blue; $H$ - green; $K$ -
  red.  The patterns broadly split into a constant offset component,
  and occasional random jumps in level believed to be caused by DC
  pedestal level offsets in the derived flatfield frames.  The numeric
  label on each panel refers to the detector with \#1 SW, \#2 SE, \#3
  NE, \#4 NW as seen on the sky.}
\label{figdetzpt}
\end{figure}

The detector zeropoint level corrections show some interesting
effects when data from five WFCAM semesters are examined (see
Fig.~\ref{figdetzpt}). The small random jumps in the temporal
sequences are most likely due to detector-level DC offsets caused by
flatfield pedestal effects (these can also be seen in the sky
corrections but in this case they are generally additive and do not
impact object photometry).  Additionally, $\approx \pm$1\,per cent, more
constant trends are seen in the offsets.  These are correlated within
a passband but are different between bands. 

We wondered if small QE variations between the detectors could give
rise to slightly different colour equations, required to relate the
natural detector+filter+telescope passband to the 2MASS standards.
The overall passband gain differences between the detectors are
corrected at the flatfielding stage but because the twilight sky used
in the flats is generally a different colour from the majority of
objects, residual colour equation differences between the detectors
could be manifest as a small zeropoint offset. Analysis of WFCAM
images shows that the twilight sky changes colour rather rapidly as
the sky darkens, ranging from fairly neutral ($J-K \sim 0.5$), through
extremely blue ($J-K \sim -2.0$), to the rather red dark sky ($J-K
\sim 2.5$). Modelling of the twilight sky and typical calibration
stars (using a range of synthetic spectra as input) suggest that the
detectors would have to exhibit large differences in QE, at the 10 per
cent level, across individual passbands to explain the observed
effect. Hewett et al. (2006) found that the QE of a typical Rockwell
Hawaii-II detector increases approximately linearly with wavelength,
by 8 per cent over the entire WFCAM 0.8--2.2 micron range. However we
currently lack measurements of the actual WFCAM detector QEs. For the
time being, the cause of the detector-to-detector photometric offsets
remains an open question.

\subsubsection{Residual spatial systematics and their possible causes}\label{ressys}

After removing the detector-level differences, generic recurring
patterns are visible in the final spatial correction, even across
different passbands.  This strongly suggests a common underlying cause
and the three most likely contributors are: 

\begin{enumerate}

\item spatially dependent PSF corrections -- the residual maps are
made from aperture-corrected 2\,arcsec diameter flux measurements.

\item low-level non-linearity in the detectors;

\item scattered light in the camera which would negate
the implicit assumption of, on average, uniform illumination of the
field;

\end{enumerate}

We have ruled out spatial PSF distortions as a significant contributor
by comparing aperture-corrected 4\,arcsec diameter fluxes with the
2\,arcsec diameter measures.  If we exclude observations made in the
first 4 months of WFCAM operations, when adjustments to the focal
plane geometry were still being made, the average systematic
difference between 2 and 4\,arcsec diameter stellar fluxes is
negligible ($<$1\,per cent) over the entire array.

\subsubsection{Residual non-linearities?}\label{spatialtests}

The spatial magnitude residuals are generally larger toward the edges
of the chips. To test the derived lookup tables, we repeated the LAS
overlap analysis (Section~\ref{repeatlas}), but this time after
correcting the source photometry. To apply the corrections we used
bilinear interpolation to compute the offset for each source in the
standard coordinate system described above. The correction is applied
additively.

\begin{figure}
\begin{center}
  \includegraphics[width=8.5cm,viewport=22 00 400
  520,clip]{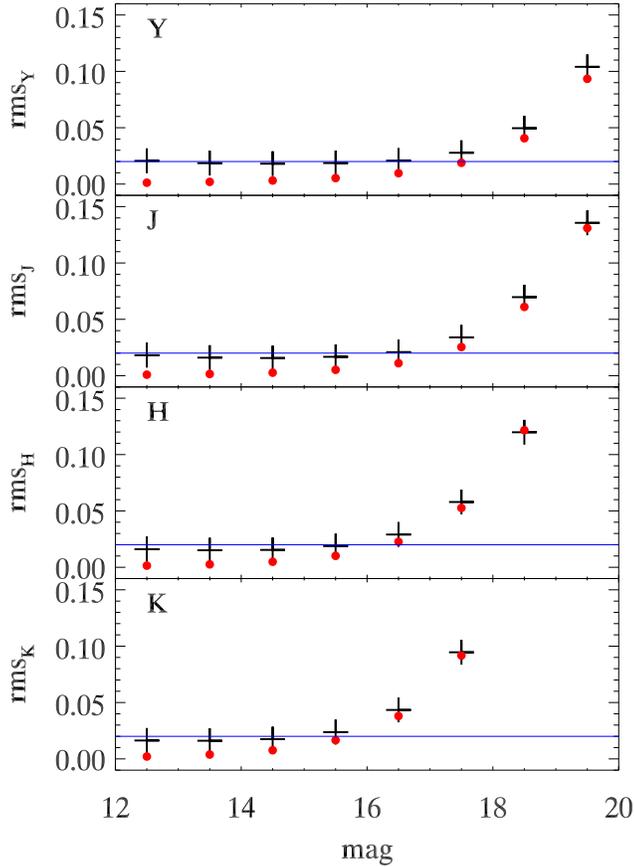}
  \caption{$rms$ diagram for repeat observations of sources in the DR3
    dataset for the LAS (selected to be stellar in both observations)
    for the $Y$$J$$H$$K$ filters (top to bottom), after correcting for
    the spatial systematics as described in the text. Data are for
    photometric conditions only. Black crosses are 1.48
    $\times$ the MAD (median absolute deviation) of the data in
    magnitude slices, i.e. equivalent to the Gaussian sigma of the
    magnitude bin. Red circles are the median pipeline errors for
    sources in the bin as reported in the WSA for DR3. The
    blue horizontal lines at 0.02 magnitudes represent the target 
    2\,per cent photometric reliability.}
\label{rms_las_cor}
\end{center}
\end{figure}

The resulting $rms$-diagram (Fig.~\ref{rms_las_cor}) shows a
significant improvement compared to the version without spatial
correction (Fig.~\ref{lasdr3repeat}). The median values of $rms$
for bright stars (taking the mag=13.5 bin in all filters), in each
passband are typically 0.002-0.004 magnitudes lower
(Table~\ref{tabrms}).

\begin{figure}
\begin{center}
  \includegraphics[width=8.0cm,viewport=0 5 480 470,clip]{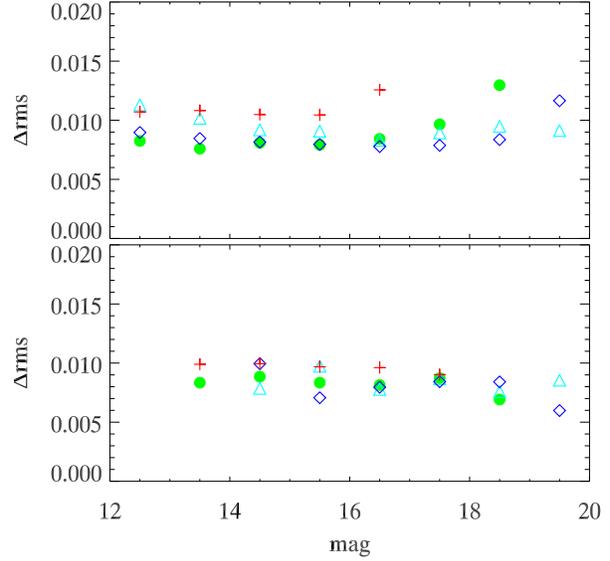}
  \caption{$\Delta rms$ is equal to $\sqrt(rms^2-rms_{\rm spatial}^2)$
    and is plotted against magnitude for the $Y$ (cyan triangle), $J$
    (blue diamond), $H$ (green filled circle) and $K$ (red plus)
    filters. In the upper panel we plot the data for objects
    classified as stellar, while the lower panel is for extended
    sources. $rms$ is the $rms$ for photometry in the DR3 release of
    the LAS (as described in Fig.~\ref{lasdr3repeat}). $rms_{\rm
      spatial}$ is the $rms$ for the same data but with photometry
    corrected for spatial systematics as described in the text.  }
\label{figcomprms}
\end{center}
\end{figure}

We can try to gain some insight into the cause of the spatial
systematics by looking at the dependence of the correction on
magnitude. This would show up any residual non-linearity effects for
example. In Fig.~\ref{figcomprms} we plot the quadrature difference
between the spatially corrected and non-corrected $rms$ as a function
of magnitude, i.e. the error contribution from the spatial
systematics. We show the diagram for stars as well as galaxies, as any
non-linearity effects would be less pronounced in spatially extended
sources. Fig.~\ref{figcomprms} shows that the spatial systematics
amount to a correction of about 1\,per cent in all filters with no 
dependence on magnitude. We conclude
that the WFCAM system shows no significant residual non-linearity.
Our conclusion is consistent with the analysis of Irwin et al. (2008)
who show WFCAM is linear to at least 1\,per cent.

\subsubsection{Scattered light as a cause of the spatial systematics?}

\begin{figure}
\begin{center}
\includegraphics[width=6.5cm,angle=270]{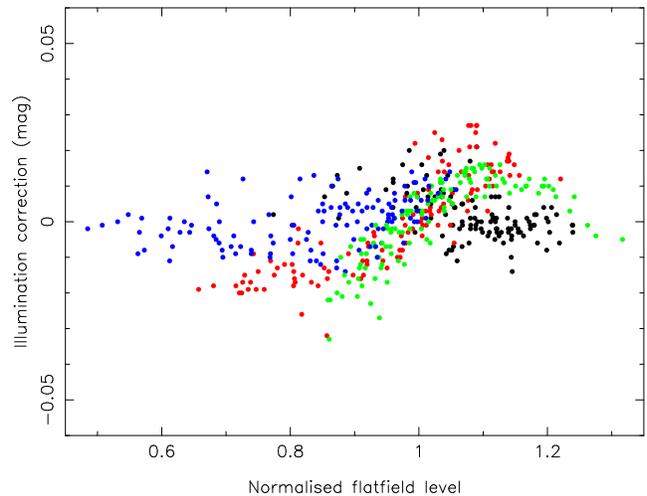}
\caption{The correlation between normalised flatfield level and the
  spatial magnitude residuals for $J$-band data from 2005
  September. Detectors \#1 (black) and \#4 (blue) show no strong
  correlation, however detectors \#2 (red) and \#3 show a well-defined
  correlation.}
\label{illumnonlin}
\end{center}
\end{figure}

Fig.~\ref{illumnonlin} shows the correlation between normalised
flatfield level and the spatial magnitude residuals for $J$-band data
analysed from 2005 September.  The detectors with the largest and
smallest flatfield variations (\#1 and \#4) show no effects at the
$\simlt$1\,per cent level; whereas detectors \#2 and \#3 show a clear
correlation between residual magnitude spatial systematics and
flatfield level. The spatial variation in the flatfield intensity is
dominated by changes in sensitivity across the detectors (and
vignetting within the optical train) and reaches a factor two for the
worst detector (Irwin et al. 2008). The spatial systematics seen in
the stacked 2MASS photometric residuals represent a 1\,per cent error
contribution, and it seems most likely to arise from (as yet
unquantified) scattered light within the WFCAM system, probably
present in all observations.

\subsection{Repeat observations of standard fields}

Standard star fields are observed every night with WFCAM, and provide
another means of testing the repeatability of the photometric
calibration. For this analysis we made use of data calibrated for
UKIDSS DR3. Twenty-seven standard reference fields were selected and,
for each field, reference stars were chosen to have a minimum of 100
detected counts in a 2 arcsecond diameter aperture. Fields typically
contain between 500 and 10,000 suitable reference stars depending on
Galactic latitude. The reference stars were then matched against
subsequent observations of the same fields taken in good conditions
(i.e. the frames pass the following criteria: seeing $\leq 1.52$
arcseconds, airmass $\leq 1.8$, NIGHTZRR $\leq 0.05$, MAGZRR $\leq
0.05$, median source ellipticity $< 0.2$). On average, the reference
fields have between 20 (in the $Z$-band) and 50 (in the
$K$-band) observations which match our criteria. The smaller
number of selected $Z$-band observations is due to the larger
error in the frame zeropoints (Section~\ref{method}).

For the repeat observations of the standard fields, the same 2MASS
calibrators are used in each observation, thus we should expect to see
an improvement in $rms$. The actual median values for $rms_{\rm STD}$
(for the 13.5 magnitude bin) range from 0.011--0.015 magnitudes for
bright stars, which are 0.004--0.006 magnitudes lower than the
spatially corrected LAS $rms_{\rm spatial}$.

\subsection{Calibration performance for data taken in non-photometric
  conditions}\label{nonphot}

\begin{figure}
\begin{center}
  \includegraphics[width=8.5cm,viewport=00 10 530 520,clip]{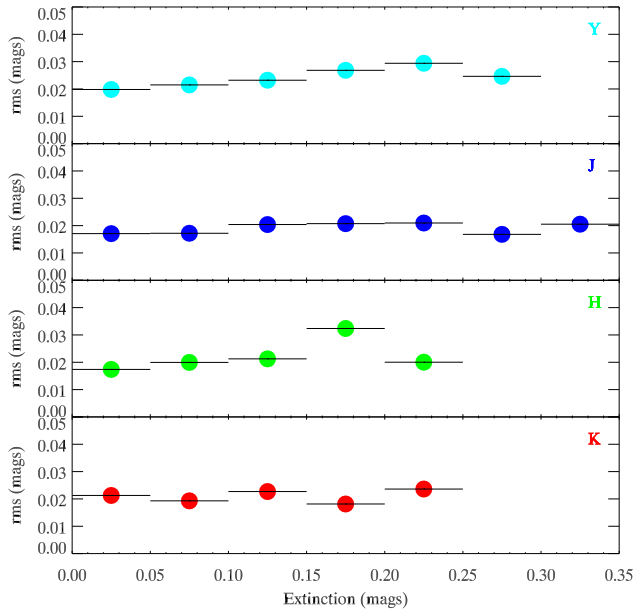}
  \caption{$rms$ as a function of sky extinction for stars with
    magitudes in the range 12--15. The sample is drawn from the LAS
    DR3 and the $rms$ values for each bin are computed from the
    magnitude differences between repeat observations of the same
    star, see Fig.~\ref{lasdr3repeat} for details. At least one of the
  frames in each pair must have an extinction value in the range
  plotted (except for the best weather bins, which must have both
  frames taken with a ZP within 0.05 mags of nominal). The data are
  plotted by filter, $YJHK$ from top to bottom.}
\label{wfcamweather}
\end{center}
\end{figure}

The measured zeropoint for each frame depends directly on the
transmission of the atmosphere. The presence of clouds leads to a
reduction in the derived zeropoint. In Fig.~\ref{wfcamweather} we
investigate the $rms$ for stars measured in different conditions to
investigate whether there is a specific sky transparency beyond which
the photometric conditions become unacceptable. We again compute the
$rms$ for magnitude differences between repeat measurements, but this
time over a larger magnitude range, m=12--15 in all filters (it's
clear from Fig.~\ref{lasdr3repeat} that the $rms$ in this magnitude
range is largely dominated by systematic errors over photon
counting). The data are plotted as a function of sky extinction as
measured by the difference of the frame zeropoint from the median
zeropoint for that filter for at least one of the frames (except for
the 'clear' sky case, where both frames must have an extinction $\leq
0.05$mag.

Fig.~\ref{wfcamweather} demonstrates that, even in non-photometric
conditions, where 10--20\,per cent of the stellar flux is lost to
cloud, the calibration is still very good, and still meets the 2\,per
cent calibration requirements.

\subsection{Photometric repeatability: summary}

In Table~\ref{tabrms} we tabulate the $rms$ at mag=13.5, for each
filter, for the different samples considered in the sections
above. Examining the first three columns, the largest calibration
improvement comes from the computation of a per-detector zeropoint, as
included in the change to WFCAM processing for DR3 from DR2. The
overlap analysis of the LAS DR3 data shows that the calibration
requirements for WFCAM are met, and that the photometry is repeatable
to $\simeq$2\,per cent, even for sources which are close to the edge
of the field of view, where residual systematics are
larger. Investigation of these systematics suggests that they are most
likely caused by scattered light present in the WFCAM flatfields,
which contributes about 1\,per cent to the overall photometric error
budget at large off-axis angles. Application of our corrections for
these systematic residuals reduces the WFCAM photometric error to only
$\simeq$1.5\,per cent in $J$, $H$ and $K$.

Repeat measurements of standard fields, where the sources are uniformly
distributed across the field of view, and the 2MASS calibrators do not
change yield a photometric repeatability at the level of 1--1.5\,per cent.

Even observations taken through thin cloud (where transmission is $>$
80\,per cent) can be calibrated to $\simeq$2\,per cent using 2MASS
stellar sources as secondary standards.

\begin{table}
\centering
\caption{$rms$ at mag=13.5 for the WFCAM filters derived from repeat
  observations of sources in the LAS and standard star fields.}
{\scriptsize
\begin{tabular}{lrrrrr}\hline
      & \multicolumn{4}{c}{LAS}        & STD  \\
Filter& $rms_{\rm DR2}$ & $rms_{\rm DR3}$ & $rms_{\rm
  spatial}$ & $rms_{\rm cloudy}$ & $rms_{\rm STD}$\\
\hline                                                              
  $Z_{\rm w}$   &   -       &  -     & -     & -     & 0.015 \\
  $Y_{\rm w}$   &   0.023   &  0.021 & 0.019 & 0.021 & 0.013 \\
  $J_{\rm w}$   &   0.021   &  0.018 & 0.016 & 0.022 & 0.012 \\
  $H_{\rm w}$   &   0.022   &  0.017 & 0.015 & 0.021 & 0.011 \\
  $K_{\rm w}$   &   0.023   &  0.020 & 0.016 & 0.020 & 0.011 \\
\hline
\end{tabular}
}
\label{tabrms}
\end{table}

\section{Comparison between WFCAM and 2MASS photometric
  systems}\label{2mcomparison}

The colour equations determined for regions of low reddening are
listed below. They were derived from fits to selected data taken in
the first year of WFCAM observations (Irwin et al. 2008). The 0.03
magnitude offset for the $H$-band is discussed below. The $Y$-band
offset was not included prior to DR3, and is discussed in
Section~\ref{yoffset}. In this Section we rederive the 2MASS--WFCAM
tranformations, and investigate their behaviour in regions of low and
high Galactic latitude. We establish the reliability of the photometry
for fields containing relatively small numbers of 2MASS
calibrators. We also quantify the effects of Galactic extinction on
the calibration, and derive additional corrections for
Equations~\ref{eqzj}--\ref{eqkk}, listed in
Equations~\ref{eqzpz}--\ref{eqzpk}.  For DR2 onwards, it is this full
set of equations that are applied to the 2MASS secondary standards to
enable derivation of the photometric zeropoint.

\begin{equation}
Z_{\rm w} = J_{2} + 0.950 (J_2-H_2)
\label{eqzj}
\end{equation}
\begin{equation}
Y_{\rm w} = J_{2} + 0.500 (J_2-H_2) + 0.080
\label{eqyj}
\end{equation}
\begin{equation}
J_{\rm w} = J_{2} - 0.065 (J_2-H_2)
\label{eqjj}
\end{equation}
\begin{equation}
H_{\rm w} = H_{2} + 0.070 (J_2-H_2) - 0.030
\label{eqhh}
\end{equation}
\begin{equation}
K_{\rm w} = K_{2} + 0.010 (J_2-K_2)
\label{eqkk}
\end{equation}

\subsection{The 2MASS colour-terms}

\begin{figure}
\includegraphics[width=8.5cm,viewport= 55 10 465 730,clip]{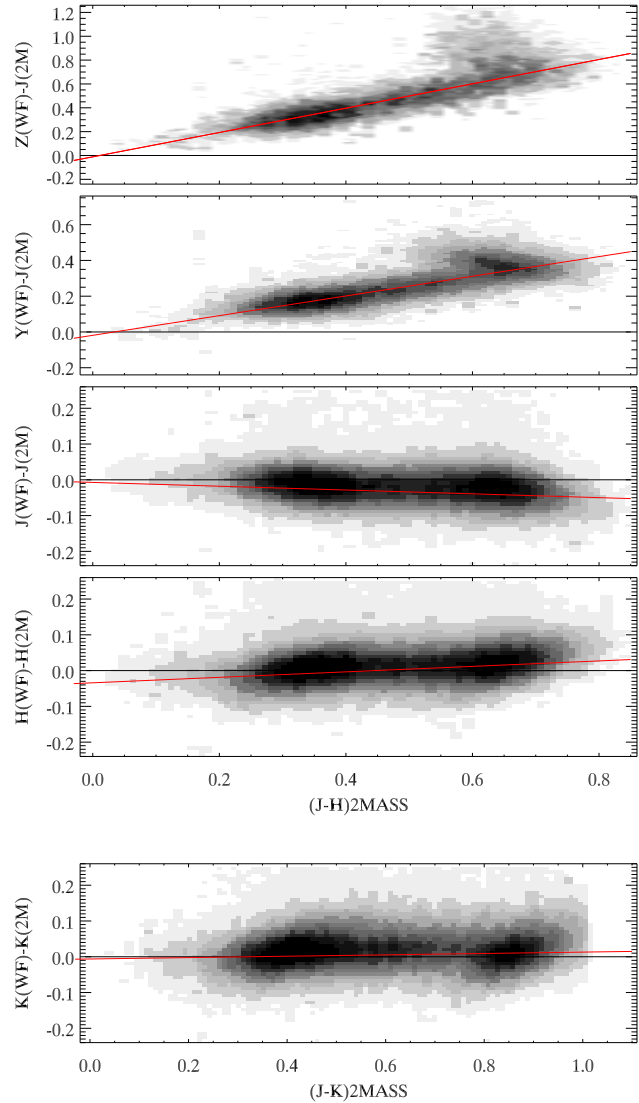}
\begin{center}
  \caption{A Hess diagram showing differences between WFCAM and 2MASS
    photometry as a function of 2MASS ($J_2-H_2$ or $J_2-K_2$) colour
    for data taken in the second half of 2005. The data have been
    selected such that $E(B-V)' \le 0.2$ and $J_2-K_2 \le 2.0$ to
    avoid the most significant effects of reddening. The best fit
    linear regressions to the unbinned data are overplotted.}
\label{2masscolours}
\end{center}
\end{figure}

Fig.~\ref{2masscolours} shows data derived from observations taken in
the second half of 2005. We plot the difference between the calibrated
WFCAM photometry and the 2MASS photometry for some 200,000
measurements across five passbands. We use these data to rederive the
colour-terms to compare to those in
Equations~\ref{eqzj}--\ref{eqkk}. The data are selected to avoid
regions of high interstellar reddening, $E(B-V)'$\footnote{Throughout
  the paper we correct the $E(B-V)$ from Schlegel et al. (1998)
  according to the prescription given by Bonifacio, Monai \& Beers
  2000, hence the use of the `prime' on the formula.}

The data are fit with a linear relation (with 3$\sigma$ clipping),
including only stars with colours in the range $0.0 \le J_2-K_2 \le
1.0$. The slopes of the fits for are shown in
Fig.~\ref{2masscolours} and can be seen to be in good
agreement with Equations ~\ref{eqzj}~to~\ref{eqkk}.

\begin{table}
\centering
\caption{Table of best fit colour terms for WFCAM--2MASS colour as a
  function of $J_2-H_2$ (or $J_2-K_2$ in the case
  of the K filter) as shown in Fig.~\ref{2masscolours}, split by
  region (see text for details). The
  first column shows the best fit colour terms for data taken in
  the second half of 2005 (semester 05B) and can be
  directly compared to the colour equations listed in
  Equations~\ref{eqzj} to \ref{eqkk}.}
\begin{tabular}{rrrrr}\hline
& Semester & \multicolumn{3}{c}{Region} \\
Colour & 05B & A & B & C \\
\hline
$Z_{\rm w}-J_2$  &  1.02  &  1.04 &  0.92 &  1.01 \\
$Y_{\rm w}-J_2$  &  0.55  &  0.51 &  0.55 &  0.56 \\
$J_{\rm w}-J_2$  & -0.05  & -0.04 & -0.06 & -0.05 \\
$H_{\rm w}-H_2$  &  0.08  &  0.04 &  0.07 &  0.07 \\
$K_{\rm w}-K_2$  &  0.02  &  0.01 &  0.01 &  0.01 \\
\hline
\end{tabular}
\label{fits}
\end{table}

For the $J$$H$$K$ passbands, the linear conversions work
surprisingly well, given the large spread in
Figure~\ref{2masscolours}. For the $H$-band, an additional
3\,per cent offset is required to bring the WFCAM photometry onto the
same system as the UKIRT faint standards (as originally published by
Hawarden et al. 2001).  The $H$-band offset is discussed in
more detail in Section~\ref{ukirtfs}.

For the $Z$ and $Y$ passbands, significant (and indeed degenerate)
residuals are seen for stars with $J_2-H_2 \sim$ 0.5---0.7. This is
most easily understood by appealing to the two-colour $J-H$ vs. $H-K$
diagram for dwarf stars, e.g. Fig.~2 in Cruz et al. (2003). Stars with
spectral types later than M0 become progressively bluer in $J-H$, but
redder in $Z-J$ and $Y-J$. Thus the same colour transformation is
applied to stars with the same $J-H$ colours but very different
temperatures and therefore $Z-J$ colours, resulting in the structures
seen in Fig.~\ref{2masscolours}.

The formal errors on all the fits are extremely small due to the large
numbers of sources present in each sample, and are not particularly
helpful when considering that they do not account for systematic
effects (e.g. population). Below we consider two issues: (1) how
robust is the determination of zeropoint assuming the colour equations
are correct for randomly selected small samples of stars, matched to
the example of the Large Area Survey?; (2) are the colour equations
significantly different in regions of high and low Galactic latitude?

\subsection{Robustness of zeropoint determination}

For the determination of the zeropoint of a typical high-latitude
field, the measured offset between the WFCAM instrumental magnitudes
and the 2MASS photometry will not be affected by these
colour-dependent residuals, so long as a robust (i.e. median) offset
is measured. Clearly the {\it average} offset in the top two panels in
Fig.~\ref{2masscolours} would be affected by the non-linearity in the
colour-transformations. If a WFCAM field were dominated by extremely
red stars, or an unusual population of objects (for example a large
and rich Globular Cluster), then a further colour-dependent zeropoint
correction may be necessary. We have yet to identify any fields that
suffer from this effect.  For example, examining zero-point variation
as a function of Galactic latitude and longitude reveals no compelling
evidence for any variation significanlty larger than 1 per cent.  We
anticipate returning to this issue in the future as more data are
collected over larger areas of sky.

The robustness of the calibration of WFCAM is most simply tested using
a Monte-Carlo approach. We took a `worst-case' scenario, comprising a
high-latitude field which contains relatively few 2MASS stars. For
each trial we selected 50 stars at random (from the dataset used
above), and derived the median offset between the 2MASS magnitudes
corrected to the WFCAM system (using Equations~\ref{eqzj}--\ref{eqkk})
and the WFCAM calibrated photometry. We performed 10,000 trials, and
summarise the results in Table~\ref{montecarlo}. This method of
calibration is found to be robust for all filters, with a standard
deviation of 1\,per cent or lower in all filters, except the $Z$-band,
for which the error is closer to 2\,per cent.

\begin{table}
\centering
\caption{Average Zeropoint offsets and their standard deviations for a
  comparison 
  between the 2MASS and WFCAM photometry for 
  10000 trials with randomly chosen samples of 50 stars}
\begin{tabular}{lrr}\hline
Filter & Mean ZP Offset & $\sigma_{\rm offset}$ \\
\hline
$Z_{\rm w}$  & -0.006  & 0.017\\
$Y_{\rm w}$  & -0.001  & 0.011\\
$J_{\rm w}$  &  0.003  & 0.007\\
$H_{\rm w}$  &  0.002  & 0.008\\
$K_{\rm w}$  &  0.003  & 0.010\\
\hline
\end{tabular}
\label{montecarlo}
\end{table}


\subsubsection{Colour equations at low and high Galactic latitude}

We defined three regions to compare: A ($0<l<100$, $|b|<10$), B
($0<l<100$, $|b|>20$), C ($180<l<250$, $|b|>15$), where $l$ and $b$
are Galactic longitude and latitude respectively. Region A thus
contains a fairly large sample close to the Galactic plane. As with
the sample examined above, we fit for the colour transformations
between the 2MASS and WFCAM filters. The results of the fitting are
listed in Table~\ref{fits}. There appear to be some small differences
between the regions, particularly in the Z-passband for region
B. However, bootstrap error estimates find an error in the slope of
$\pm0.02$ for regions B and C in this filter, so we do not believe
these differences to be significant (above 2$\sigma$). Note that the
colour term for a pointing would need to be adjusted by 0.04 to change
the zeropoint by 2\,per cent for stars of median $J_2-H_2=0.5$.

\subsection{The Effects of Galactic extinction}\label{sec:galext}

\begin{figure}
\begin{center}
\includegraphics[width=9.3cm]{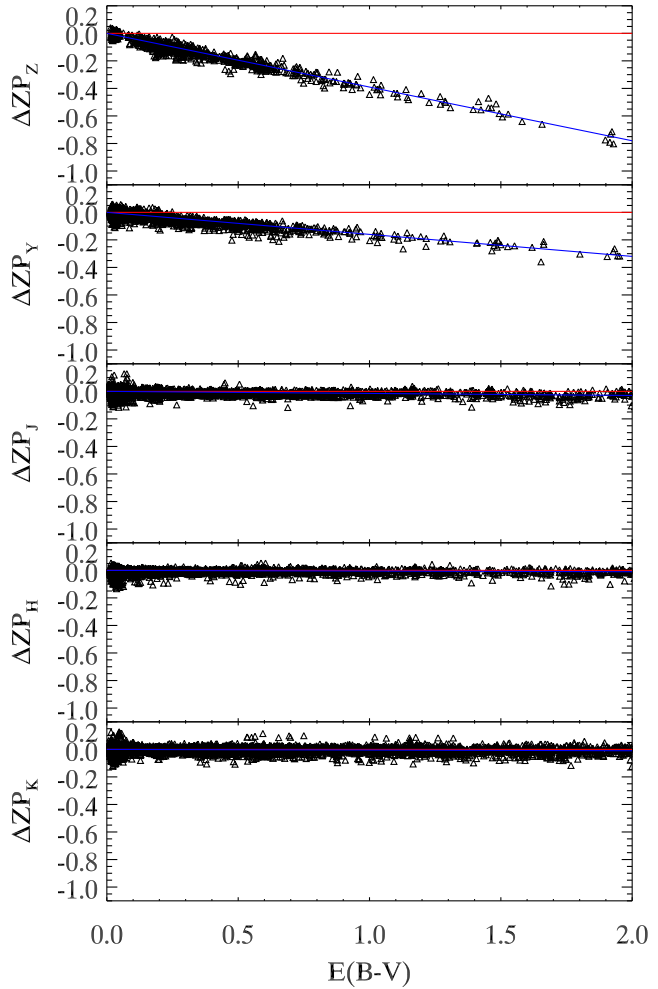}
\caption{Frame zeropoint minus nightly zeropoint (uncorrected for
  reddening) versus $E(B-V)'$ for the WFCAM broadband filters. Only
  data from photometric nights are included. Data points which lie
  more than 0.1 magnitudes away from the fit are clipped. The blue
  line is the best fit relation to the extinction relation, while the
  red line is constant $\rm \Delta ZP=0.0$.}
\label{extinction}
\end{center}
\end{figure}

We have attempted to mitigate against the effects of interstellar
reddening on the WFCAM calibration by employing a colour restriction
to the stars used in the calibration (applied after extinction
correction). At regions of low Galactic latitude, where reddening is
high, the spectra of distant stars are strongly affected by
intervening dust. The population mix of the WFCAM calibrators will
change as intrinsically blue stars are moved into the applied colour
selection, and redder stars are moved beyond the colour bounds. Giant
stars become significantly more common. Differences between the WFCAM
and 2MASS filter profiles become important, and the colour equations
used to transform the 2MASS magnitudes onto the WFCAM system could
begin to break down. For the $J$$H$$K$ filters, for a mixed stellar
population, these differences are small, and we expect the effect on
the calibration to be similarly small. For the $Z$ and $Y$ filters,
where we are extrapolating from the $J_2$ magnitudes using the
$J_2-K_2$ colour then we may expect to see significant reddening
dependent offsets in the calibration.

The following analysis makes use of data calibrated according to the
prescription for DR1, which made no attempt to correct for
reddening. In Fig.~\ref{extinction} we plot the difference between the
nightly averaged zeropoint and the zeropoints measured for individual
fields taken on that same night, versus $E(B-V)'$ (corrected according
to Bonifacio, Monai \& Beers 2000) for the $Z$$Y$$J$$H$$K$ filters
using data taken on photometric nights between 2005 April and 2007
May. The $E(B-V)'$ value for each frame is computed as the average
$E(B-V)'$ of all stars contributing to the calibration of the
detector. Data points which lie $>0.1$ magnitudes outside the best fit
slopes are clipped.

A significant, $E(B-V)'$ dependent, correction is clearly required for
the $Z$ and $Y$ filters. The correction required for the $J$, $H$ and
$K$ passbands is more than an order of magnitude smaller, for example
$\simeq$3\,per cent in $J$ for a change in $\Delta E(B-V)' =
2.0$. Fits to these data are shown in Fig.~\ref{extinction} and listed
in Equations~ref{eqzpz}--\ref{eqzpk}. The colour equations employed
for UKIDSS DR2 onwards have been modified to include these additional
$E(B-V)'$ terms.

\begin{figure}
\begin{center}
\includegraphics[width=8cm]{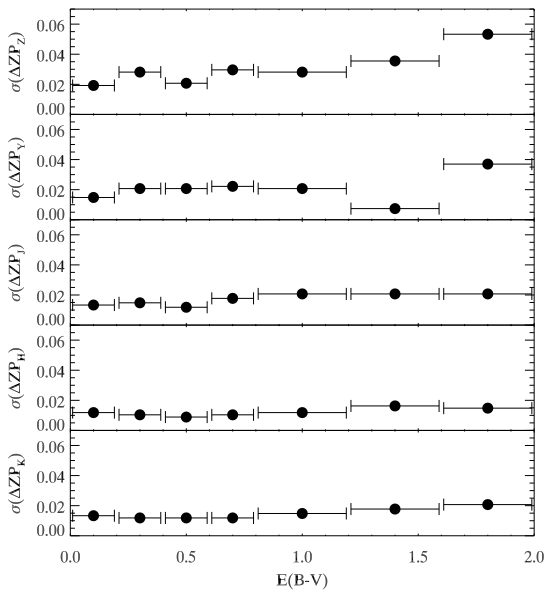}
\caption{The standard deviation of $\Delta$ZP (computed from the MAD)
  around the best fit relations described in Equations~\ref{eqzpz} to
  \ref{eqzpk} as a function of $E(B-V)'$. The data are binned in
  $E(B-V)'$ as indicated by the horizontal bars.}
\label{madzp}
\end{center}
\end{figure}

\begin{equation}
ZP'(Z) = ZP(Z) - 0.390 E(B-V)'
\label{eqzpz}
\end{equation}
\begin{equation}
ZP'(Y) = ZP(Y) - 0.160 E(B-V)'
\end{equation}
\begin{equation}
ZP'(J) = ZP(J) - 0.015 E(B-V)'
\end{equation}
\begin{equation}
ZP'(H) = ZP(H) - 0.005 E(B-V)'
\end{equation}
\begin{equation}
ZP'(K) = ZP(K) - 0.005 E(B-V)'
\label{eqzpk}
\end{equation}

There exists a paucity of measurements at high reddening for the Z and
Y passbands, because they are not included in the Galactic Plane
Survey. Nevertheless, we investigate the scatter in the measured
zeropoints as a function of $E(B-V)'$ and try to estimate at what
point the reddening becomes sufficiently large that the photometric
calibration is compromised. This is attempted in Fig.~\ref{madzp}
where we plot the standard deviation (computed from the median
absolute deviation) of the data around the fits (Equations~\ref{eqzpz}
to \ref{eqzpk}) as a function of $E(B-V)'$.  $\sigma_{\rm ZP}$ is
somewhat larger for the $Z$ and $Y$ bands at all
values of $E(B-V)'$, as expected. For the $J$$H$$K$ bands,
the scatter in the measured ZP is seen to be very slowly rising as a
function of $E(B-V)'$. Our conclusion is that the 2MASS-based
calibration reaches the WFCAM goal for $E(B-V)' \le 2.0$. For the
$Z$-band, the WFCAM calibration goals are not met for
$E(B-V)'>0.2$, while for the $Y$-band, the WFCAM calibration appears
to be robust up to $E(B-V)'=1.5$.

In summary, and more qualitatively, the zeropoints, and therefore the
calibration, derived for highly reddened fields appear to be robust in
the $J$-, $H$- and $K$-bands, but should be treated with some caution
in the $Z$- and $Y$-bands.

\section{The effects of the overestimation of flux for faint
  sources in 2MASS}\label{bias}

\begin{figure*}
\begin{center}
\includegraphics[width=17cm,viewport=38 10 620 250,clip]{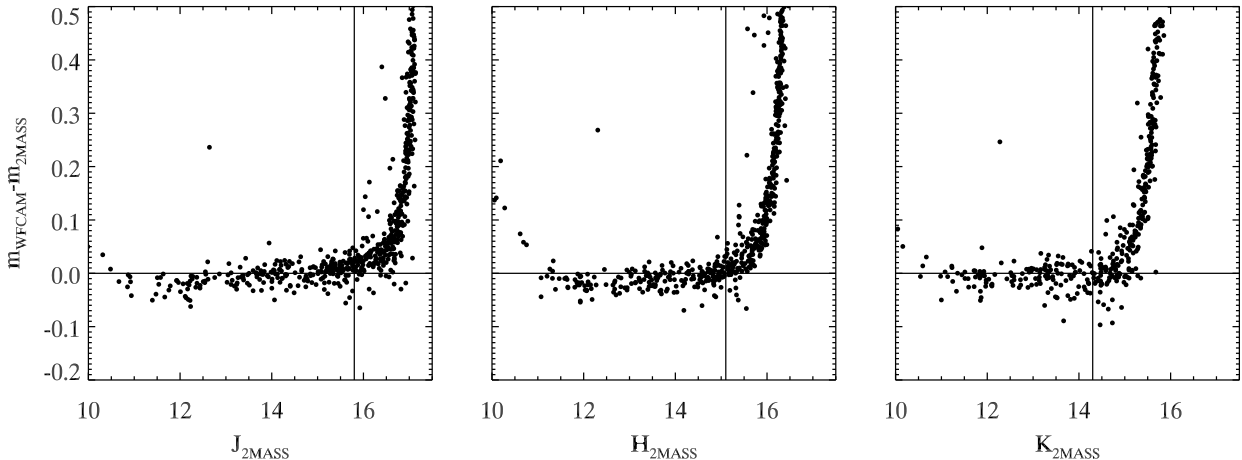}
\caption{Differences between measured WFCAM photometry for calibration
  fields and the same sources in the 2MASS Calibration Merged Point
  Source Information Tables
  (www.ipac.caltech.edu/2mass/releases/allsky/doc/seca6\_1.html)
  plotted as a function of 2MASS magnitude for all filters,
  illustrating the effects of Eddington Bias. The 10$\sigma$ magnitude
  thresholds for calibration star selection are shown as a vertical
  line.}
\label{2mmvswf}
\end{center}
\end{figure*}

By calibrating WFCAM from a shallower survey, we open ourselves up to
the possibility of a photometric bias affecting the results. All
measured fluxes in a survey are subject to uncertainties, and close
to the detection threshold, where the uncertainties are large, this
effect will result in a bias, whereby sources with negative
fluctuations will not be detected, while those with positive
fluctuations will be pushed to brighter magnitudes. Thus, on average,
sources near to the detection threshold of a survey will have
overestimated fluxes. This is generally known as the Eddington Bias
(Eddington 1940). 

In order to investigate the impact of Eddington Bias on the WFCAM
calibration, we make use of the 2MASS Calibration Merged Point Source
Information
Tables\footnote{http://www.ipac.caltech.edu/2mass/releases/allsky/doc/\\
seca6\_1.html}. These
tables contain the results of merging catalogues from the individual
2MASS calibration scans to give average measured photometry and
astrometry for all sources observed multiple times. We found 7 of the
2MASS calibration scans overlapped with WFCAM standard fields with at
least 5 observations (though most had $>100)$.

In Fig.~\ref{2mmvswf} we compare magnitudes from the 2MASS merged
tables with observations of the same sources in WFCAM. Note that the
WFCAM magnitudes combine measurements of at least five observations for
all objects. The WFCAM calibration is derived from 2MASS point
sources with a signal-to-noise ratio $\ge 10$. As can be seen in
Fig.~\ref{2mmvswf}, the Eddington Bias at these levels is $\le 2$\,per
cent in all filters, indicating that the error introduced into the
WFCAM calibration is negligible. We also note for the $J$-band, there
is a possible trend between magnitude and the magnitude difference, suggestive
of a small non-linearity. The effect is less pronounced at $H$ and
probably non-existent at $K$.

\section{The WFCAM $Z$$Y$$J$$H$$K$ system}\label{secoffsets}

WFCAM photometry is on a Vega system (see Hewett et al. 2006). The
photometric zeropoints for all the WFCAM filters for each observed
frame are derived by measuring the offsets between the 2MASS
calibrators (now converted to the WFCAM system) and the observed
stars, as described in Section~\ref{method} using the colour equations
from Section~\ref{2mcomparison}. For the $Z$ and $Y$ passbands in
particular, there are no ready calibrators available, and there is an
underlying assumption that the 2MASS colours can be linearly
extrapolated into the WFCAM system. In this section we investigate the
WFCAM photometric system in more detail using colour-colour diagrams,
and in particular concentrating on photometrically selected A0 stars.

\subsection{Sources in the SDSS}

\begin{figure}
\begin{center}
\includegraphics[width=8.5cm,viewport=35 50 450 410,clip]{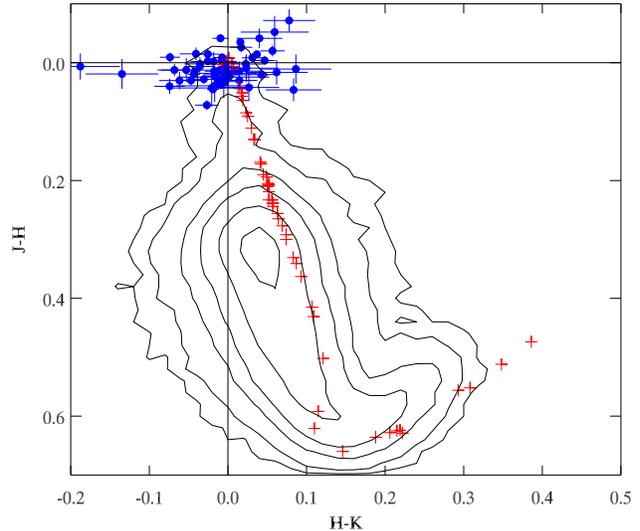}
\caption{The $J-H$ vs $H-K$ diagram for stellar sources in the UKIDSS
  LAS with counterparts in SDSS. Objects are selected to be stellar in
  all WFCAM filters and an overall stellar SDSS
  classification. Measured photometric errors are $\le 0.05$ 
  magnitudes in each WFCAM passband. Sources with SDSS colours close
  to zero ($u-g, g-r, r-i, i-z$ all in the range -0.1 to 0.1) are
  plotted with errors. Red crosses are synthetic photometry from
  Hewett et al. (2006) based on the BPGS atlas (see text for details).}
\label{lassdss_jhk}
\end{center}
\end{figure}

We define a sample of stellar sources detected in both the UKIDSS
Large Area Survey (LAS) and SDSS. In addition we highlight stars with
SDSS photometry consistent with a spectral type A0, i.e. having Vega
colours close to zero ($u-g, g-r, r-i, i-z$ all in the range -0.1 to
0.1\footnote{We convert all SDSS photometry onto a VEGA system,
  applying the offsets from AB magnitudes listed in Hewett et
  al. 2006}). This sample is used to investigate the WFCAM photometry
and check that the infrared colours are also close to zero. The $J-H$
vs $H-K$ diagram is shown in Fig.~\ref{lassdss_jhk}. The median WFCAM
colours for the candidate A0 stars are $J-H=0.014 \pm 0.024$ and
$H-K=-0.008 \pm 0.037$ and are consistent with zero (with errors
derived as the robust standard deviation of the data, $\sigma=1.48
\times MAD$). In conclusion the stellar sequence is not offset in the
JHK diagram. Synthetic WFCAM photometry is presented by Hewett et
al. (2006) derived from the Bruzual-Persson-Gunn-Stryker (BPGS)
spectrophotometric atlas. In Figure~\ref{lassdss_jhk} we plot the
synthetic stellar sequence and we can see that there is a significant
offset from the data. At this point it's not clear whether the offset
should be attributed to either (a) the initial calibration of the BPGS
spectrophotometry, (b) the synthetic colours generated by Hewett et
al. (2006), or (c) systematic shifts present in the WFCAM calibration.

\begin{figure}
\begin{center}
\includegraphics[width=9.3cm,viewport=5 30 520 818,clip]{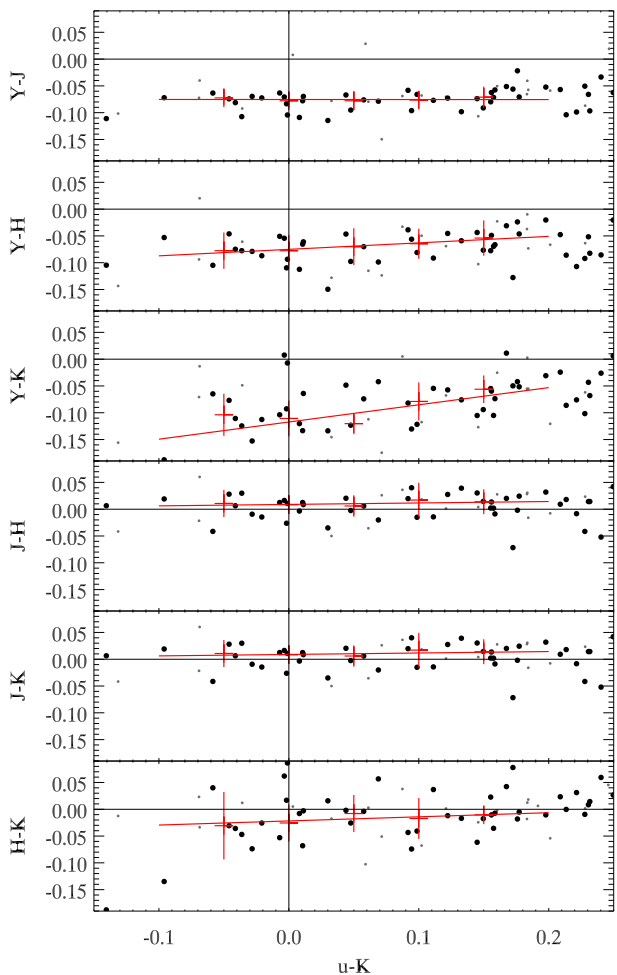}
\caption{WFCAM colours for stellar sources in common with SDSS. Large
  symbols are selected to be blue in SDSS ($u-g, g-r, r-i, i-z$ all in
  the range -0.1 to 0.1). Error bars are the standard deviations on
  binned median values. The best fit straight lines are also drawn.}
\label{figoffsets}
\end{center}
\end{figure}

We can explore this in a little more detail for each of the LAS
filters in turn. We use the $u-K$ colour to extend the baseline on the
x-axis, and plot all WFCAM colour combinations in
Fig.~\ref{figoffsets} for the $Y$$J$$H$$K$ filters. Offsets are
measured from straight line fits to the blue stars. The slopes of the
fits are very shallow, and thus any systematic error in the $u'$- or
$K$-band photometry will not significantly affect the measured
offset. At $u-K=0.0$, we find $\Delta_{YJ}=-0.075 \pm 0.007$,
$\Delta_{YH}=-0.075 \pm 0.018$, $\Delta_{YK}=-0.117 \pm 0.018$,
$\Delta_{JH}=-0.009 \pm 0.012$, $\Delta_{JH}=-0.009 \pm 0.012$,
$\Delta_{HK}=-0.022 \pm 0.024$, indicating an appreciable offset for
the $Y$-band, and little or no offset in the $J$$H$$K$ filters.

\subsection{The $Y$-band}\label{yoffset}

\begin{figure}
\begin{center}
\includegraphics[width=9.3cm]{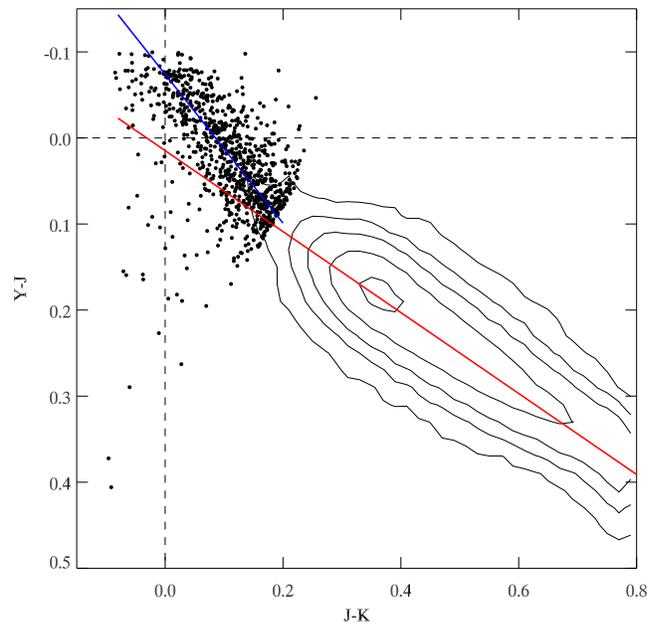}
\caption{The DR2 $Y-J$ vs $J-K$ diagram for stellar sources common to
  the UKIDSS LAS and SDSS surveys. Objects are selected to be stellar
  in all WFCAM filters with a measured photometric error $\le 0.05$
  magnitudes in each passband. SDSS sources are selected to have a
  stellar classifcation and a photometric error $\le 0.1$ in all
  filters. Sources in regions of significant reddening, $E(B-V)'>0.2$
  are excluded from the plot and from analysis. So-called `blue stars'
  are plotted with a larger symbol. The best-fit straight line is
  shown in blue. An additional linear fit to the remaining stars is
  shown in red extending down to $J-K=0.8$.}
\label{figlassdss_yjjk}
\end{center}
\end{figure}

At the start of WFCAM observations, there was no reliable calibration
for $Y$, and as a consequence an estimate of the zeropoint was made
for the early data releases. To examine the $Y$-band calibration in
more detail, we plot the $Y-J$ vs $J-K$ diagram in
Fig.~\ref{figlassdss_yjjk}. The blue star sequence does not pass
through (0,0), but is significantly shifted to the blue in $Y-J$. We
attribute the displacement to the offset in the $Y$-band calibration
in DR2.  A maximum likelihood straight-line fit to the blue stars
(including photometric errors, see Appendix) yields the intercept in
$Y-J$, $\Delta_{Y}=-0.074 \pm 0.005$.

Relaxing the source selection such that there is no requirement
for a counterpart in SDSS yields a slightly larger sample. From
the LAS dataset alone we measure $\Delta_{Y}=-0.080 \pm 0.005$.

Fig.~\ref{figlassdss_yjjk} shows that there is a
non-linearity in the transformation between the $J-K$ and $Y-J$
colours for blue stars. Hence the $Y$-band offset is a natural
consequence of adopting a linear extrapolation based on redder stars.

\subsection{The $Z$-band}

\begin{figure}
\begin{center}
\includegraphics[width=8cm]{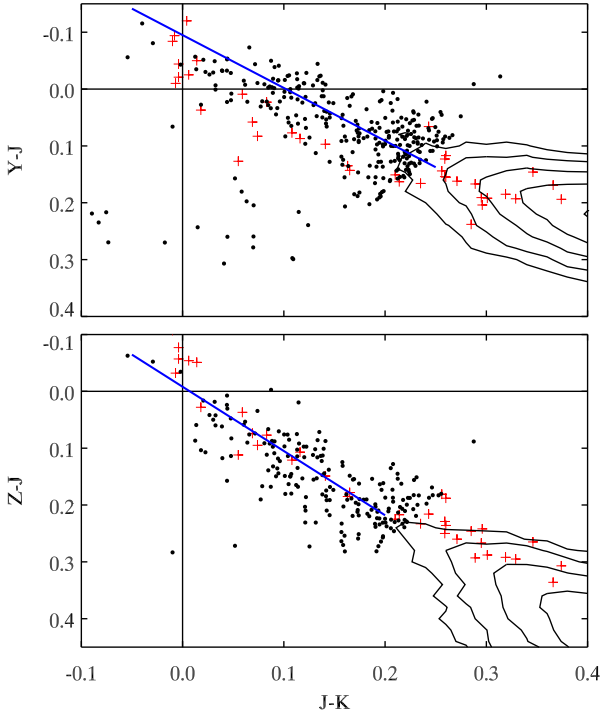}
\caption{Colour-colour diagrams for UKIDSS GCS photometry of point
  sources. Red crosses are synthetic photometry from Hewett et
  al. (2006).}
\label{gcszyjjk}
\end{center}
\end{figure}

The only UKIDSS survey with significant $Z$-band coverage is the
Galactic Clusters Survey (GCS), with essentially no overlap with SDSS
at present. In Fig.~\ref{gcszyjjk} we plot $Y-J$ and $Z-J$ against
$J-K$, for stellar sources detected in the GCS to: (i) confirm the 
$Y$-band offset discussed above, (ii) check the validity of using the 
GCS dataset to measure the offset, and (iii) investigate a possible offset 
in the $Z$-band.

The sample of blue stars is rather smaller than derived from the LAS
sample, but the offset for the $Y$-band is reproduced with
$\Delta_{YJ}=-0.095 \pm 0.015$. For the $Z$-band we measure an
intercept of $\Delta_{ZJ}=-0.008 \pm 0.015$, i.e. consistent with
zero.

\subsection{Summary of offsets}

In summary, an offset to the WFCAM $Y$-band zeropoint needs to be
applied to bring it onto the Vega photometric system. We take the
weighted mean of the various values derived above with respect to the
WFCAM $J$-band and the SDSS $i$-band, and find $\Delta_{Y}=-0.078 \pm
0.010$. Consequently an offset of 0.080 has been applied for UKIDSS
DR3 and subsequently as shown in Equation~\ref{eqyj}.

For the $ZJHK$-bands, the data are consistent with no offset at the
$\pm 2$\,per cent level.

\section{Comparison with published $J$$H$$K$ photometry in the Mauna
  Kea Observatories photometric system}\label{ukirtfs}

Every night that WFCAM is observing, a number of standard fields are
measured, typically every two hours (in the first year of operations
this was an hourly procedure). The majority of the standard fields
include stars selected from the list of UKIRT faint standards
published in Hawarden et
al. (2001)\footnote{http://www.jach.hawaii.edu/UKIRT/astronomy/calib/phot\_cal/}.
These fields were initially chosen to act as the primary source of
photometric calibration for WFCAM, however, subsequent analysis
(presented in this paper) has demonstrated that the 2MASS-based
calibration offers significant advantages. Observations of standard
stars currently continue to be performed at the telescope, to ensure
that a future calibration can be derived independently from 2MASS
should the need arise. The standard observations also enable
monitoring of the WFCAM system by the reliable strategy of repeatedly
looking at the same stars.

Recently, Leggett et al. (2006) have revisited the UKIRT faint
standards, and present new measurements using the UKIRT Fast Track
Imager (UFTI). They find small offsets (at the few\,per cent level)
with respect to the older data, as well as some evidence of
magnitude-dependent non-linearity, and recommend adopting the newer
photometry.

\begin{figure}
\begin{center}
\includegraphics[width=8.5cm,viewport=20 28 500 700,clip]{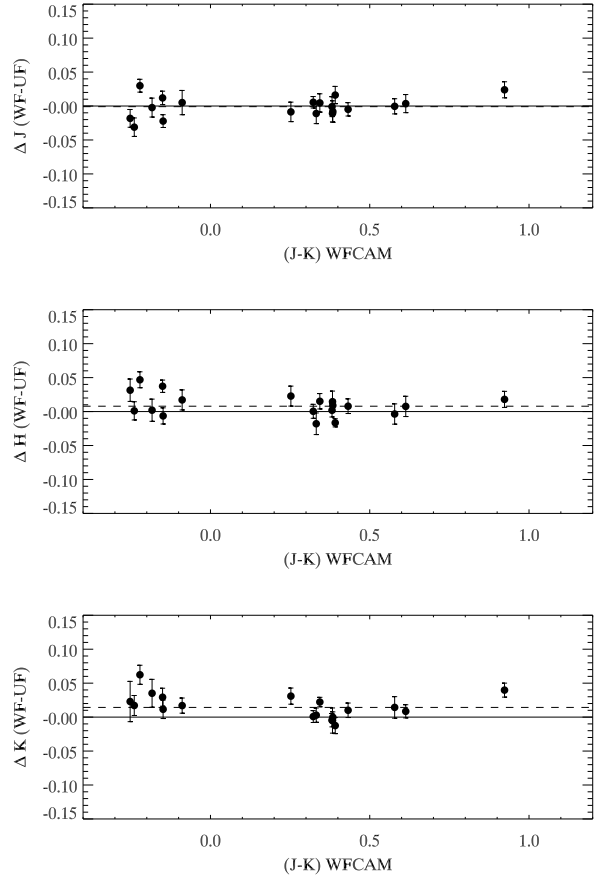}
\caption{Median $\Delta$mag (WFCAM minus UFTI photometry) plotted as a
  function of $J-K$ (UFTI) for the $J$, $H$ and $K$ filters (top to
  bottom). The dashed lines show the median measured offsets between
  the WFCAM and UFTI photometry, discussed in the text.}
\label{figfs}
\end{center}
\end{figure}

In Fig.~\ref{figfs} we compare the DR3 WFCAM calibration with the
recalibrated UFTI photometry of faint standard stars measured in the
Mauna Kea Observatories system (Leggett et al. 2006).

WFCAM measurements were selected from observations taken on
photometric nights during good conditions. There are typically 50--100
measurements per star in each filter. The data values plotted are the
median $\rm m_{\rm w}-m_{UFTI}$ for the $J$$H$$K$ filters, and error
bars are the standard deviation calculated as 1.48$\times$ the median
absolute deviation (MAD) of the WFCAM measurements. Only stars passing
the following criteria are plotted and analysed: (i) they show no
variability ($rms_{JHK}<=0.03$); (ii) they are fainter than
$J$$H$$K$=11 to avoid possible saturation/non-linearity effects. Thus
the sample is rather small, and does not include any objects with
extreme colours.

Most stars are not variable at the 1--2\,per cent level, but we note a
few exceptions: FS117 (b216-b9), FS143 (Ser-EC86), GSPC P259-C, which
all show evidence for variability at $\ge 5$\,per cent within the
timeframe of the observations (2005 April--2007 May). As in Leggett et
al. (2006), we find FS144 (Ser-EC84) may be variable at the 3\,per cent
level. However, we find no evidence to suggest that FS116 (b216-b7) is variable
at $\ge 1$\,per cent.

Over the limited colour range considered in Fig.~\ref{figfs}, we do
not see evidence for a colour term between the $J_{\rm w}$ and
$J_{\rm UFTI}$ photometry. However, a larger sample of stars with both
UFTI and WFCAM photometry, spanning a broader colour range, is
required before a definitive statement can be made. The median
measured offsets between the WFCAM and UFTI photometry are:
$\delta_{J}=0.005$, $\delta_{H}=0.015$, $\delta_{K}=0.017$. WFCAM-UFTI
is positive in the three filters, i.e. the WFCAM measurements are
slightly fainter. Leggett et al. (2006) also measured small offsets
between the WFCAM and UFTI systems; and find a different offset in the
$H$-band, with $H_{\rm w}-H_{\rm UFTI} \sim -0.015$. We note that
the Leggett et al. (2006) measurements are based on an earlier WFCAM
calibration, and use a sample of stars without such a restrictive
magnitude cut and comprising many fewer repeat observations.

During the WFCAM Science Verification phase (2005 April), a decision
was taken to anchor the WFCAM calibration to the Hawarden et
al. (2001) UKIRT photometry. An offset of 0.03 magnitudes was found
between the 2MASS-based WFCAM calibration and the UKIRT system, thus
requiring a correction to the colour equation (Equation~\ref{eqhh})
for the $H$-band. When making this decision, we were also aware that
Cohen et al. (2003) argued for small offsets to be applied to the
2MASS $H_2$ and $K_2$ bands to bring them into agreement with their
space-based infrared spectrophotometric calibration system (the offset
required for $J_2$ is negligible). The suggested offsets, to be applied
to the published 2MASS photometry, were $-0.019 \pm 0.007$ ($H_2$), and
$+0.017 \pm 0.005$ ($K_2$).

At present, in the absence of a larger sample of observations and
more-accurate photometry, we conclude that the WFCAM photometric
system is a Vega system (to within 2\,per cent) and is in agreement
with all three of the Hawarden et al. (2001), Cohen et al. (2003) and
Leggett et al. (2006) systems to within 2\,per cent. For the time
being we choose to retain the 3\,per cent offset in the conversion
between the 2MASS and WFCAM $H$-band photometry.

%

\section{Empirical determination of the throughput of
  WFCAM}\label{throughput}

\subsection{Gain}\label{gain}


Measurements of the readout noise and gain from the 2005 April SV data
are given in Table~\ref{tabgain}. The readout noise was estimated from the
difference between successive single exposure CDS 5\,s dark frames after
running the decurtaining algorithm to remove systematic
artifacts. Note that the average dark current is generally negligible
and dominated by reset anomaly variations. The gain is the average of
gains measured at three background levels 23k, 14k and 5k; the overall
variation in measured gain was at the ~0.05 level with no clear trends
with background.

\begin{table}
\centering
\caption{Measured dark current, uncorrected gain
  and noise (in electrons and ADU) for the four WFCAM detectors. The
  gain is an overestimate, as described in the text. The corrected
  average gain for WFCAM is 4.31.}
\begin{tabular}{l|rrrr}\hline
Det       & Dark curr. & Noise (ADU) &Uncorr. gain  & Noise (e-) \\\hline
     \#1  &        -2.4  &        3.8      & 4.84  &     18.4\\
     \#2  &         3.7  &        4.2      & 4.87  &     20.5\\
     \#3  &        -9.0  &        4.2      & 5.80  &     24.4\\
     \#4  &        49.1  &        4.4      & 5.17  &     22.7\\\hline
\end{tabular}
\label{tabgain}
\end{table}

Whilst measuring the gain from the dome flat sequences, an estimate of
the inter-pixel capacitance via a robust measure of the noise
covariance matrix was also made. A sum of the noise covariance matrix
close to 0,0 gives a value of 1.20 which implies that the total
reduction in directly measured noise variance (i.e. measured using a
conventional method) is therefore also 1.20. We therefore predict a
~20\,per cent overestimate of the gain, and hence a ~20\,per cent
overestimate of the QE coefficients. Such an overestimate is as
expected for the Rockwell Hawaii-II devices. The average gain of the
WFCAM detectors is therefore 4.31.

\subsection{Throughput}

The total throughput of the system is relatively easy to measure (but
it is much harder to quantify where in the system the actual losses
are occurring). For the calculation we have assumed the following: The
effective area of the UKIRT primary mirror is 10.5m$^2$ (i.e. outer
diameter 3.802m with an inner diameter of 1.028m). No attempt has been
made to allow for the shadowing of the primary by the secondary, nor
for any obscuration caused by the forward mounted camera itself.  The
spectrum for Vega is the same used by Hewett et al. (2006), originally
from Bohlin \& Gilliland (2004).  The passband transmissions are from
Hewett et al. (2006) but renormalized to give the relative throughput
as a function of wavelength, rather than absolute transmission
(i.e. the peak value is 1.0).

Thus Table~\ref{tabthrough} gives the estimated number of photons that
would be incident on the primary mirror, assuming no atmosphere,
multiplied by the relative transmission of the filter in each
band. The values in table~\ref{tabthrough} allow the calculation of the 
zeropoint of the system in each filter assuming no losses due to the 
atmosphere, telescope and the instrument. Comparison with the median 
measured zeropoints for DR1  then give us the throughput for 
WFCAM+UKIRT+atmosphere.

\begin{table}
\centering
\caption{Estimate of WFCAM throughput}
\begin{tabular}{l|rrrrr}
Filt & N$_{\nu}$(0$^{\rm m}$) & N$_{\rm C}$(0$^{\rm m}$) & ZP$_{100}$ & ZP
(M) & TP$_{\rm M}$\\\hline
$Z_{\rm w}$ & 3.7e10 & 8.5e9 & 24.95 & 22.77 & 15\% \\      
$Y_{\rm w}$ & 3.1e10 & 7.3e9 & 24.77 & 22.78 & 18\% \\      
$J_{\rm w}$ & 2.9e10 & 6.8e9 & 24.70 & 22.97 & 23\% \\      
$H_{\rm w}$ & 2.8e10 & 6.6e9 & 24.66 & 23.22 & 29\% \\      
$K_{\rm w}$ & 1.6e10 & 3.6e9 & 24.02 & 22.55 & 29\% \\\hline
\end{tabular}
\caption{Where N$_{\nu}$ is the number of photons reaching the
  detector in each passband for a zeroth magnitude star, N$_{\rm c}$
  is the equivalent number of counts in the detector and ZP$_{100}$ is
    the corresponding predicted zeropoint for each filter if there
    were no losses in the system. ZP$_{\rm M}$ is the actual measured
    median zeropoint for WFCAM in each filter, and TP$_{\rm M}$ is
    therefore the adjusted throughput of the system compared to a
    perfect instrument and atmosphere.}
\label{tabthrough}
\end{table}



\section{Conclusions}

\begin{enumerate}

\item The CASU pipeline photometric calibration of WFCAM data, using
  2MASS sources within each WFCAM pointing, has achieved a photometric
  accuracy better than 2\,per cent for the UKIDSS Large Area Survey
  released in DR3 (Data Release 3) and later, based on repeat
  measurements.

\item By stacking spatially binned WFCAM data on monthly timescales we
  have shown that photometric calibration residuals are present at the
  1\,per cent level. We have derived and tested a method for removing
  these residuals, and show that the photometric calibration for the
  WFCAM $J$$H$$K$ filters can reach an accuracy of 1.5\,per cent. We
  attribute these residuals to scattered light in the WFCAM twilight
  flatfield frames. These corrections have been applied to the UKIDSS
  DR4.

\item Even observations taken through thin cloud (where transmission
  is $>$ 80\,per cent) can be calibrated to $\sim$ 2\,per cent using
  2MASS calibrators.

\item Monte-Carlo sampling the 2MASS stars observed by WFCAM, suggests
  that as long as $\sim$ 50 calibrators fall within the field-of-view,
  then the WFCAM calibration is robust to 1\,per cent in $Y$$J$$H$$K$
  and 2\,per cent in $Z$. 99\,per cent of the sky has more than 50 2MASS
  calibrators in a single WFCAM pointing.

\item The WFCAM calibration incorporates colour restrictions on
  secondary standards and modified colour equations to mitigate against
  the effects of Galactic extinction. Towards regions of significant
  reddening, the calibration meets the 2\,per cent
  requirement for $E(B-V)' \leq 2.0$ for the $J$$H$$K$ bands, but
  should be treated with caution above $E(B-V)' = 0.2$ for the $Z$
  band and to a lesser extent for the $Y$ band.

\item We find that the calibration of WFCAM from the shallower 2MASS
  survey is free from the effects of Eddington Bias, thanks to a
  careful choice of a signal-to-noise cut (SNR$=10$) for the 2MASS
  sources.

\item We identified a small offset between the WFCAM $Y$ photometry
  and an idealized Vega photometric system in the second UKIDSS Data
  Release (DR2). A correction of $\Delta_Y=-0.08$ magnitudes has now
  been included in the WFCAM calibration and applied from UKIDSS DR3,
  in the sense that the $Y$ band sources are now 0.08 magnitues
  fainter than in previous releases.

\item We find that the WFCAM photometric system is within 0.02
  magnitudes of the UKIRT MKO system (Leggett et al. 2006) for the
  $J$$H$$K$ filters, and is Vega-like to within 2\,per cent
  (i.e. stars selected to have A0 colours in SDSS have WFCAM colours
  $<=0.02$ across all passbands).

\end{enumerate}

\section*{Acknowledgments}

Our thanks go to the staff of UKATC and UKIRT for building and
operating WFCAM, and to the staff at WFAU for operating the WFCAM
Science Archive which has been used extensively. We also thank Richard
McMahon, Sandy Leggett, Nigel Hambly and the anonymous referee for
their input to this paper.

\appendix

\section{Structural Analysis}

\begin{figure}
\label{fig:a1}
\begin{center}
{\includegraphics[width=0.7\hsize,angle=270]{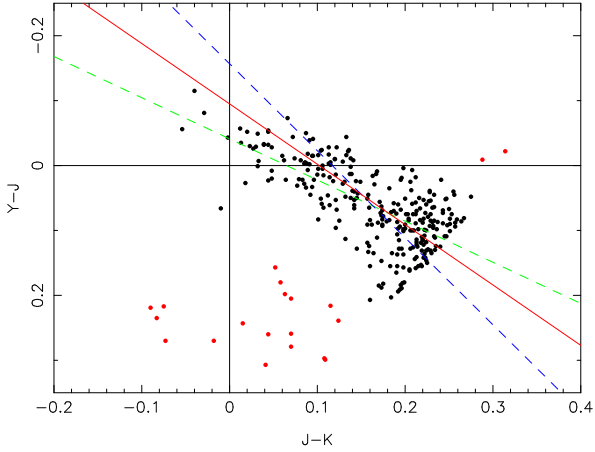}}
\caption{Locating the offset in the $Y$-band calibration for WFCAM using 
standard least-squares regression (green and blue lines) compared to the
correct Maximum likelihood technique (red line).  Red filled points
are excluded from all the fits.}
\end{center}
\end{figure}

We use the blue end of the distribution of points culled from the GCS to
illustrate the problem of finding an accurate estimate for the $Y$-band
offset. In this case, since the WFCAM $J$,$K$-bands are already on the Vega
system, this is equivalent to finding the intercept on the $Y-J$ axis when 
the $J-K$ colour is zero.  However, with significant errors on both axes, 
and possibly also an intrinsic spread in the stellar locus, conventional 
least-squares curve fitting methods can give seriously biased results.  
This is illustrated in Fig.~\ref{fig:a1} where the obvious outliers 
(red points) have been excluded from the fits.  In this example we have 
(arbitrarily) defined the x-axis to be the $J-K$ measures and the y-axis to 
be the $Y-J$ measures.  The green dashed line shows the result of a standard 
variance-weighted least-squares regression of $y$ on $x$, while the blue 
dashed line shows the same for a regression of $x$ on $y$.  The values of
the supposedly equivalent slopes and intercepts are significantly different
and the clear bias in both fits is simply the result of ignoring the duality 
of the error distribution.  We note that an often-used shortcut for this
type of situation is simply to ``average'' the results from the two 
regressions and use that for the solution.  As it happens, if the magnitude 
of the errors on both axes are similar this often gives a result close to
the optimum value.

The correct procedure in this type of case is to explicitly recognise
the more complex error distribution involved by rephrasing the modelling 
problem in terms of the unobserved ``true'' parameters using the method
of maximum likelihood.  This technique is also known as structural analysis
(\eg Kendall \& Stuart 1979).  

In general terms the underlying model is of the form $Y_i = f(X_i \mid \theta)$,
whereas what we observe are not these values but rather the quantities
$x_i = X_i + \delta x_i $ and $y_i = Y_i \  + \delta y_i $ where $\delta x_i$
and $\delta y_i$ are independent errors with variance 
$ \sigma^2_{x_i} $\ , \ $\sigma^2_{y_i} $ which may include intrinsic
spread in the model as well as observational and calibration errors.  
If these errors are Gaussian distributed then the likelihood $L$ of observing 
the $N$ independent pairs of values $x_i,y_i$ is given by 

\begin{equation}
 L = \prod_{i=1}^N \ P(x_i,y_i \mid \theta) 
\end{equation}

where $\theta$ represent the model parameters; and the log-likelihood is
therefore given by

\begin{equation}
 ln(L) = -N\ ln(2\pi) - {1 \over 2}\ \sum_i \ ln(\sigma^2_{x_i}
        . \sigma^2_{y_i}) - {1 \over 2} \sum_i \left (
 {\delta x_i^2 \over \sigma^2_{x_i}} + {\delta y_i^2 \over 
 \sigma^2_{y_i} } \right ) 
\end{equation}

For unknown errors the problem is insoluble.  For known errors

\begin{equation}
  ln(L) = const - {1 \over 2} \sum_i {[x_i - X_i]^2 \over \sigma^2_{x_i}}
         + {[y_i - Y_i]^2 \over \sigma^2_{y_i}}
\end{equation}

and the solution effectively involves solving for $X_i$ and $\theta_j$ \ie 
$N+m$ unknowns, often using a mixture of a direct parameter search on $ln(L)$
in conjunction with the following differential constraints

\begin{equation}
\label{diff}
  {\partial ln(L) \over \partial X_i} = {(x_i - X_i) \over \sigma^2_{x_i}}
   + {\partial f \over \partial X_i}\ \left [{y_i - f(X_i \mid \theta) \over
   \sigma^2_{y_i}} \right ] = 0 
\end{equation}

\begin{equation}
  {\partial ln(L) \over \partial \theta_j} = 
  \sum_i \ {\partial f \over \partial \theta_j} \ \left [{ y_i - f(X_i \mid 
  \theta)
  \over  \sigma^2_{y_i}} \right ] = 0 
\end{equation}

\begin{figure}
\label{fig:a2}
\begin{center}
{\includegraphics[width=0.7\hsize,angle=270]{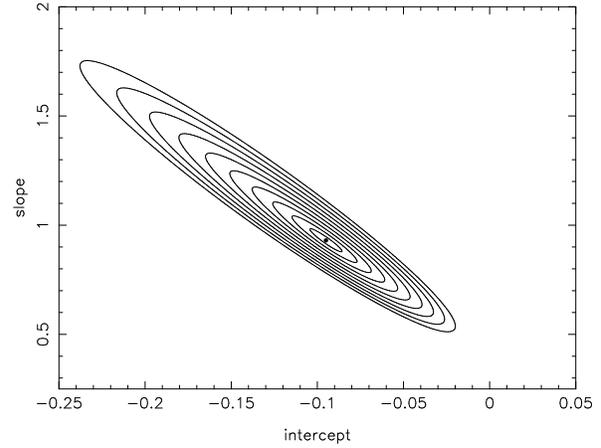}}
\caption{Log-likelihood contours starting at 1-$\sigma$ and spaced by intervals 
of 1-$\sigma$ for the maximum likelihood model fit shown for the data presented
in figure~\ref{fig:a1}}
\end{center}
\end{figure}

We illustrate the use of this technique to find the intercept using the 
straight line model defined by $Y_i = a X_i +b$.  In this case the differential
constraint on $X_i$ (equation~\ref{diff}) simplifies to 

\begin{equation}
  {x_i - X_i \over \sigma^2_{x_i}} + {a(y_i - aX_i -b) \over \sigma^2_{y_i}} = 0 
\end{equation}

which implies for a given parameter pair $(a,b)$ that $X_i$ is determined by

\begin{equation}
  X_i = { a y_i - a b + x_i \lambda_i^2 \over a^2 + \lambda_i^2}
\end{equation}

where we have defined $ \lambda_i^2 = {\sigma^2_{y_i} \over \sigma^2_{x_i}} $.

Given $X_i$ and a trial pair $(a,b)$, $Y_i$ is defined by the model and it is
then straightforward to compute the log-likehood function using a direct 
search on a suitable grid of values $(a,b)$.  The result of doing this for
the data presented in Fig.~\ref{fig:a1} is shown in the log-likelihood 
contours in Fig.~\ref{fig:a2}.

\section{SQL}

An example WFCAM Science Archive SQL query for the selection of repeat
$K$-band observations in the Large Area Survey. Note that the
\begin{tt} s1.sourceID < s2.sourceID \end{tt} qualifier is
used to ensure that each match appears only once in the output results
(and not twice as would happen otherwise).

\begin{verbatim}

SELECT s1.kAperMag3 as mag1,s2.kAperMag3 as mag2

FROM   lasSource s1,lasSource s2,
       lasMergeLog l1,lasMergeLog l2,
       lasDetection d1,lasDetection d2,
       MultiframeDetector f1,MultiframeDetector f2,
       CurrentAstrometry c1,CurrentAstrometry c2  
       lasSourceNeighbours x,

WHERE  s1.sourceID = x.masterObjID AND          
       s2.sourceID = x.slaveObjID AND
       s1.frameSetID = l1.frameSetID AND 
       s2.frameSetID = l2.frameSetID AND        
       s1.kAperMag3Err < 0.12 AND
       s2.kAperMag3Err < 0.12 AND
       s1.kclass = -1 AND 
       s2.kclass = -1 AND 
       f1.multiframeID = c1.multiframeID AND 
       f2.multiframeID = c2.multiframeID AND 
       f1.extNum = c1.extNum AND            
       f2.extNum = c2.extNum AND            
       d1.multiframeID = f1.multiframeID AND 
       d2.multiframeID = f2.multiframeID AND 
       d1.extNum = f1.extNum AND            
       d2.extNum = f2.extNum AND            
       d1.objID = s1.kObjID AND
       d2.objID = s2.kObjID AND
       l1.kmfID = f1.multiframeID AND
       l2.kmfID = f2.multiframeID AND                    
       l1.keNum = f1.extNum AND 
       l2.keNum = f2.extNum AND
       s1.frameSetID <> s2.frameSetID AND
       s1.sourceID < s2.sourceID AND 
       distanceMins < (0.3/60.0) AND            
       distanceMins IN (
              SELECT  MIN(distanceMins) 
              FROM    lasSourceNeighbours 
              WHERE   masterObjID=x.masterObjID
                       ) 

\end{verbatim}

\section{Conversion of flux into magnitudes}

The processing philosophy is to preserve the image and catalogue data
as counts, and to document all the required calibration information in
the file headers. Thus recalibration of the data requires only changes
to the headers, and these headers can be reingested into the WSA
without the need to reingest the full tables. For readers accessing
the flat files (catalogues and images) rather than the WSA database
products, we document the methods for converting the fluxes into
magnitudes and calibrating the photometry.

\begin{equation}
m=ZP-2.5log_{\rm 10}(\frac{f}{t})-A-
2.5log_{\rm 10}(\frac{f_{\rm cor}}{f})-k(\chi-1) 
\end{equation}

where $ZP$ is the zeropoint for the frame (keyword: MAGZPT in the FITS header),
$f$ is the flux within the chosen aperture (e.g. column: APER\_FLUX\_5),
$t$ is the exposure time for each combined integration (keyword: EXP\_TIME), and
$A$ is the appropriate aperture correction (e.g. keyword: APCOR5). The next
term deals with the distortion correction caused by the varying pixel
scale, where ($f_{\rm cor}/f$) is the correction and is
derived in Equation~\ref{eq:distortcor}. The final term deals with the
extinction correction, where $k$ is the extinction coefficient
(EXTINCTION) and is equal to 0.05 magnitudes/airmass in all filters,
and $\chi$ is the airmass (keywords: AMSTART, AMEND).

\section{Relevant source and image parameters}

\begin{table*}
\caption{Table of source parameters generated by the WFCAM pipeline and
  written to the FITS catalogue
  products, and an accompanying short description for each.}
\begin{tabular}{|p{1cm}|p{3.0cm}|p{11cm}|}
\hline
1 & Sequence number & Running number for ease of reference, in strict
order of image detections.\\
2 & Isophotal flux & Standard definition of summed flux within detection isophote, apart from detection filter is used to define pixel connectivity and hence which pixels to include. This helps to reduce edge effects for all isophotally derived parameters.\\
3 & X coord & Intensity-weighted isophotal centre-of-gravity in X.\\
4 & Error in X & Estimate of centroid error.\\
5 & Y coord & Intensity-weighted isophotal centre-of-gravity in Y.\\
6 & Error in Y & Estimate of centroid error.\\
7 & Gaussian sigma & Derived from the three intensity-weighted second
moments. The equivalence to a generalised elliptical Gaussian
distribution is used to derive:

Gaussian sigma = $(\sigma_a^2+\sigma_b^2)^{1/2}$\\

8 & Ellipticity & ellipticity = $1.0-\sigma_a/\sigma_b$\\
9 & Position angle & position angle = angle of ellipse major axis wrt x axis\\
10--16& Areal profile 1 

Areal profile 2

Areal profile 3

Areal profile 4

Areal profile 5

Areal profile 6

Areal profile 7
& Number of pixels above a series of threshold
levels relative to local sky. Levels are set at T, 2T, 4T, 8T ... 128T
where T is the threshold. These can be thought of as a poor
man's radial profile. For deblended, i.e. overlapping
images, only the first areal profile is computed and the rest are set
to -1.\\
17 & Areal profile 8 & For blended images this parameter is used to flag the start of the sequence of the deblended components by setting the first in the sequence to 0\\
18 & Peak height & In counts relative to local value of sky - also zeroth order aperture flux\\
19 & Error in peak height & \\
20--45 & Aperture flux 1 

Error in flux 

Aperture flux 2

Error in flux 

Aperture flux 3

Error in flux 

Aperture flux 4

Error in flux 

Aperture flux 5

Error in flux 

Aperture flux 6

Error in flux 

Aperture flux 7

Error in flux 

Aperture flux 8

Error in flux 

Aperture flux 9

Error in flux 

Aperture flux 10

Error in flux 

Aperture flux 11

Error in flux 

Aperture flux 12

Error in flux 

Aperture flux 13

Error in flux 
& These are a series of different radius
soft-edged apertures designed to adequately sample the curve-of-growth
of the majority of images and to provide fixed-sized aperture fluxes
for all images. The scale size for these apertures is selected by
defining a scale radius $\sim<$FWHM$>$ for site+instrument. In the case of
WFCAM this "core" radius (rcore) has been fixed at 1.0 arcsec for
convenience in inter-comparison with other datasets. A 1.0 arcsec
radius is equivalent to 2.5 pixels for non-interleaved data, 5.0
pixels for 2x2 interleaved data, and 7.5 pixels for 3x3 interleaved
data. In $\sim$1 arcsec seeing an rcore-radius aperture contains roughly
2/3 of the total flux of stellar images. (The rcore
parameter is user specifiable and hence is recorded in the output
catalogue FITS header.)

The aperture fluxes are sky-corrected integrals (summations) with a
soft-edge (i.e. pro-rata flux division for boundary pixels). However,
for overlapping images the fluxes are derived via simultaneously
fitted top-hat functions, to minimise the effects of crowding. Images
external to the blend are also flagged and not included in the large
radius summations.

Aperture flux 3 is recommended if a single number is required to
represent the flux for ALL images - this aperture has a radius of
rcore.

Starting with parameter 20 the radii are: (1) $1/2 \times$rcore, (2)
$1/\sqrt2 \times$rcore, (3) rcore, (4) $\sqrt2 \times$rcore, (5) $2
\times$rcore, (6) $2\sqrt2 \times$rcore, (7) $4 \times$rcore, (8)
$5 \times$rcore, (9) $6 \times$rcore, (10) $7 \times$rcore,
(11) $8 \times$rcore, (12) $10 \times$rcore, (13) $12 \times$rcore.

Note $4 \times$rcore contains $\sim$99\% of PSF flux.

The apertures beyond Aperture 7 are for generalised galaxy photometry.

Note larger apertures are all corrected for pixels from overlapping
neighbouring images.

The largest aperture has a radius 12$\times$rcore ie. $\sim$24 arcsec diameter.

The aperture fluxes can be combined with later-derived aperture
corrections for general purpose photometry and together with parameter
18 (the peak flux) give a simple curve-of-growth measurement which
forms the basis of the morphological classification scheme.\\ 
46 & Petrosian radius & $r_p$ as defined in Yasuda et al. 2001 AJ 112
1104 \\
47 & Kron radius & $r_k$ as defined in Bertin and Arnouts 1996 A\&A Supp
117 393 \\
48 & Hall radius & $r_h$ image scale radius eg. Hall \& Mackay 1984 MNRAS
210 979\\\hline
\end{tabular}
\end{table*}

\begin{table*}
\begin{tabular}{|p{1cm}|p{3.0cm}|p{11cm}|}
\hline
49 & Petrosian flux & Flux within circular aperture to $k \times r_p$; $k$ = 2\\
50 & Error in flux & \\
51 & Kron flux & Flux within circular aperture to $k \times r_k$; $k$ = 2\\
52 & Error in flux& \\
53 & Hall flux & Flux within circular aperture to $k \times r_h$; $k$ = 5;
alternative total flux\\ 
54 & Error in flux& \\
55 & Error bit flag & Bit pattern listing various processing error flags\\
56 & Sky level & Local interpolated sky level from background tracker\\
57 & Sky $rms$ & local estimate of $rms$ in sky level around image\\
58 & Child/parent & Flag for parent or part of deblended deconstruct (redundant since only deblended images are kept)\\
59-60 & RA 

DEC
& RA and Dec explicitly put in columns for overlay programs
that cannot, in general, understand astrometric solution coefficients
- note r*4 storage precision accurate only to ∼50mas. Astrometry can
be derived more precisely from WCS in header and XY in parameters 5 \&
6\\
61 & Classification & Flag indicating most probable morphological classification: eg. -1 stellar, +1 non-stellar, 0 noise, -2 borderline stellar, -9 saturated\\
62 & Statistic & An equivalent N(0,1) measure of how stellar-like an
image is, used in deriving parameter 61 in a "necessary but not
sufficient" sense. Derived mainly from the curve-of-growth of flux
using the well-defined stellar locus as a function of magnitude as a
benchmark (see Irwin et al. 1994 SPIE 5493 411 for more details).\\ \hline
\end{tabular}
\end{table*}

\begin{table*}
\caption{Relevant photometric parameters measured by the pipeline and
  written to the FITS headers. These values are computed per-detector
  and stored in the headers for each image and catalogue
  extension. The names by which the parameters are stored in the WFCAM
Science Archive tables are also given.}
\begin{tabular}{|p{2.5cm}|p{2.5cm}|p{9cm}|}
\hline
FITS keyword & WSA parameter & Description\\\hline
AMSTART & amStart & Airmass at start of observation\\
AMEND & amEnd & Airmass at end of observation\\
PIXLSIZE & pixelScale & $[$arcsec$]$ Pixel size\\
SKYLEVEL & skyLevel& $[$counts/pixel$]$ Median sky brightness\\
SKYNOISE & skyNoise & $[$counts$]$ Pixel noise at sky level\\
THRESHOL & thresholdIsoph & $[$counts$]$ Isophotal analysis threshold\\
RCORE & coreRadius & $[$pixels$]$ Core radius for default profile fit\\
SEEING & seeing & $[$pixels$]$Average stellar source FWHM\\
ELLIPTIC & avStellarEll & Average stellar ellipticity (1-b/a)\\
APCORPK & aperCorPeak & $[$magnitudes$]$ Stellar aperture correction -- peak height\\
APCOR1 & aperCor1 & $[$magnitudes$]$ Stellar aperture correction -- core/2 flux\\
APCOR2 & aperCor2 & $[$magnitudes$]$ Stellar aperture correction -- core/$\sqrt2$ flux \\
APCOR3 & aperCor3 & $[$magnitudes$]$ Stellar aperture correction -- core flux \\
APCOR4 & aperCor4 & $[$magnitudes$]$ Stellar aperture correction --
$\sqrt2 \times$ core flux \\
APCOR5 & aperCor5 & $[$magnitudes$]$ Stellar aperture correction -- $2
\times$ core flux\\
APCOR6 & aperCor6 & $[$magnitudes$]$ Stellar aperture correction --
$2\sqrt2 \times$
core flux\\
APCOR7 & aperCor7 & $[$magnitudes$]$ Stellar aperture correction -- $4
\times$ core flux \\
MAGZPT & photZPExt &  $[$magnitudes$]$ Photometric ZP for default extinction\\
MAGZRR & photZPErrExt & $[$magnitudes$]$ Photometric ZP error\\
EXTINCT & extinctionExt &$[$magnitudes$]$ Extinction coefficient\\
NUMZPT & numZPCat & Number of 2MASS standards used\\
NIGHTZPT & nightZPCat & $[$magnitudes$]$ Average photometric ZP for the filter for
the night\\
NIGHTZRR & nightZPErrCat & $[$magnitudes$]$ Photometric ZP $\sigma$ for the filter for the night\\
NIGHTNUM & nightZPNum & Number of ZPs measured for the filter for the night\\
\hline
\end{tabular}
\end{table*}

\end{document}